\documentclass[preprint,12pt]{aastex}

\begin{document}

\title{Evidence for Companion-Induced Secular Changes in the 
Turbulent Disk of a Be Star in the LMC~MACHO~Database}

\author{Mitchell F. Struble, \altaffilmark{1}}
\affil{Department of Physics and Astronomy,
	209 South 33$^{rd}$ Street,
	University of Pennsylvania, Philadelphia, PA 19104}
	
\author{Anthony Galatola, \altaffilmark{2}}
\affil{Department of Geology and Astronomy,
	West Chester University, West Chester, PA 19383}

\author{Lorenzo Faccioli, \altaffilmark{3}}
\affil{Department of Physics and Astronomy,
	209 South 33$^{rd}$ Street,
	University of Pennsylvania, Philadelphia, PA 19104}

\author{Charles Alcock, \altaffilmark{4}}
\affil{Harvard-Smithsonian Center for Astrophysics,
	60 Garden Street, MS 45,
	Cambridge, MA 02138}
	
\and

\author{Kelle Cruz \altaffilmark{5}}
\affil{Department of Astrophysics, AMNH, \\
Central Park West at 79th Street, New York, NY 10024} 

\altaffiltext{1}{\textsl{struble@physics.upenn.edu}}
\altaffiltext{2}{\textsl{agalatola@wcupa.edu}}
\altaffiltext{3}{\textsl{faccioli@student.physics.upenn.edu}}
\altaffiltext{4}{\textsl{calcock@cfa.harvard.edu}}
\altaffiltext{5}{\textsl{kelle@amnh.org}}
\begin{abstract}
The light curve of a blue variable in the MACHO LMC database (FTS ID 78.5979.72) appeared nearly unvarying for about 4 years (the quasi-flat segment) but then rapidly changed to 
become periodic with noisy minima for the remaining 4 years 
(the periodic segment); there are no antecedent indications of a 
gradual approach to this change.
Lomb Periodogram analyses indicate the presence of two distinct periods
of $\sim 61$ days and 8 days in \textit{both} the quasi-flat and the 
periodic segments. 
Minima of the periodic segment cover at least 50\% of the orbital 
period and contain spikes of light with the 8-day period; maxima do not
show this short period.  
The system typically shows maxima to be redder than minima. 
The most recent OGLE-III light curve shows only a 30-day periodicity.
The variable's $V$ and $R$ magnitudes and color are those of a Be 
star, and recent sets of near infrared spectra four days apart, secured
during the time of the OGLE-III data, show H$\alpha$ emission near and 
at a maximum, confirming its Be star characteristics.
\par
The model that best fits the photometric behavior consists of a thin 
ring-like circumstellar disk of low mass with four obscuring sectors orbiting 
the central B star in unison at the 61-day period.
The central star peers through the three equi-spaced separations 
between the four sectors producing the 8-day period.
These sectors could be dusty vortices comprised of particles larger 
than typical interstellar dust grains that dim but selectively scatter 
the central star's light, while the remainder of the disk contains 
hydrogen in emission making maxima appear redder.
\par
A companion star of lower mass in an inclined and highly eccentric orbit produces an impulsive perturbation near its periastron to
change the disk's orientation, changing eclipses from partial to complete within $\sim 10$ days.
The most recent change to a 30 day period observed in the OGLE-III 
data may be caused by obscuring sectors that have coalesced into larger ones and spread out along the disk.
\end{abstract}

\section{Introduction}
The MACHO\footnote{\url{http://www.macho.mcmaster.ca/}} database contains nearly continuous photometric coverage of $\sim 10^{5}$ variable stars over an 8-year time span \citep{cook95}.
The LMC variable star FTS (Field, Tile, Sequence) ID 78.5979.72 in the MACHO database is located at
($\alpha$, $\delta$, J2000)=(5h 17m 27.958s,
$-69^{\arcdeg} 34^{\prime} 31.75^{\prime\prime}$).
We used transformations from MACHO instrumental magnitudes to standard
Kron-Cousins $V$ and $R$ magnitudes given by K. Cook 
\citetext{private communication}:
\begin{eqnarray}
\label{eq:eq1}
V&=&V_{MACHO}+24.22-0.1804(V_{MACHO}-R_{MACHO}) \nonumber
\\
R&=&R_{MACHO}+23.98+0.1825(V_{MACHO}-R_{MACHO}).
\end{eqnarray}
The variable has a mean apparent blue magnitude $V$ of 15.877 mag, a 
mean apparent red magnitude $R$ of 15.839 mag, and a mean color
$\vr$ of 0.038 mag.
Note that these mean apparent magnitudes differ from those in 
\citet{keller02} and from the calibrated magnitudes in the plots from 
the MACHO website since we used the transformations given by Eq.
(\ref{eq:eq1}).
For the LMC, these apparent magnitudes and color are those of a star
slightly to the right of the main sequence where Be stars reside.
Assuming a distance modulus for the LMC of 18.5 \citep{benedict02},
$M_{V}=-2.62$ and $M_{R}=-2.66$ (uncorrected for reddening); these are
consistent with a B2-B3 star ($7-5.6~\mathrm{M}_{\odot}$, 
$4.9-4.3~\mathrm{R}_{\odot}$,
$T_{eff}=19300-16000~\mathrm{K}$, \citet{lyubimkov02}).
\par
The OGLE-II \footnote{\url{http://sirius.astrouw.edu.pl/\~{}ogle/}}
ID is 051728.14-693431.7, its mean $I$ band magnitude
$I=15.993$ (DIA photometry, \citep{udalski97,zebrun01}), and
the time span of the light curve overlaps the last half of
the MACHO data, with $\sim 150$ day extension past its end,
with gaps in coverage.
Observations in the $V$ and $B$ bands are much sparser over that 
same time span with mean magnitudes $V=16.037$ and $B=16.047$ (DoPHOT
photometry).
The GSC Catalog 2.2 ID is S013203120987; $F$ band and $V$ band 
magnitudes are 16.07 and 15.73 respectively (at approximately JD 
2450364, which is just past the middle of the MACHO data, at a 
maximum).  
In the 2MASS survey it is less resolved than the OGLE image, and is 
among the fainter images on the $J$, $H$, and $K_{s}$ band images.
We estimate an upper limit 2MASS $J$ band magnitude to be 16 and $H$
band magnitude to be even fainter from the available images; its image 
is indistinguishable from noise in the $K_{s}$ band.
The object is listed in neither the 2MASS nor DENIS catalogs. 
\par
The object is present in the AGAPEROS survey fields (i.e., the EROS~1 
CCD dataset of 1991, \citep{melchior00}) of red variables in the 
LMC~bar (in the field of variables AGPRS051723.55-693420.6 and 
AGPRS051733.23-693420.2)
but is not cataloged as a variable star, presumably because it was too 
blue and its amplitude too small at this time and so did not meet the 
selection criterion for variability.
The object is present on both $B$ and $V$ prints, at $~16.5$ and $16$ 
mag respectively (estimated from a comparison with cataloged variables 
nearby), of the Hodge-Wright Atlas of the LMC \citep{hodge67}.
It is not cataloged as a variable star, presumably because its 
amplitude was too small at this time and so did not meet the selection 
criterion for variability.
With an epoch of 1968 of the original plates, we estimate an upper 
limit to its proper motion of $<0.01$ arcsec yr$^{-1}$.
\par
The object has been classified as a blue variable in the LMC MACHO database and is assigned a variability "mode" of 1, characterized as a "bumper" variable.
The majority of MACHO blue variables are typically reddest at 
maximum, and, for those examined spectroscopically (about 8\% of the 
sample), 91\% exhibit variable Balmer emission characteristic of Be 
stars; the emission occurs at or near maxima of the light curves 
\citep{keller02}, typical of Be stars \citep{dachs88}, including those 
in the LMC and SMC \citep{grebel97,keller98}.
\section{Light Curves, Lomb Periodogram Analysis, and Near-IR Spectra}
\label{sec:lc}
\begin{figure} 
\begin{center}
\footnotesize
\plotone{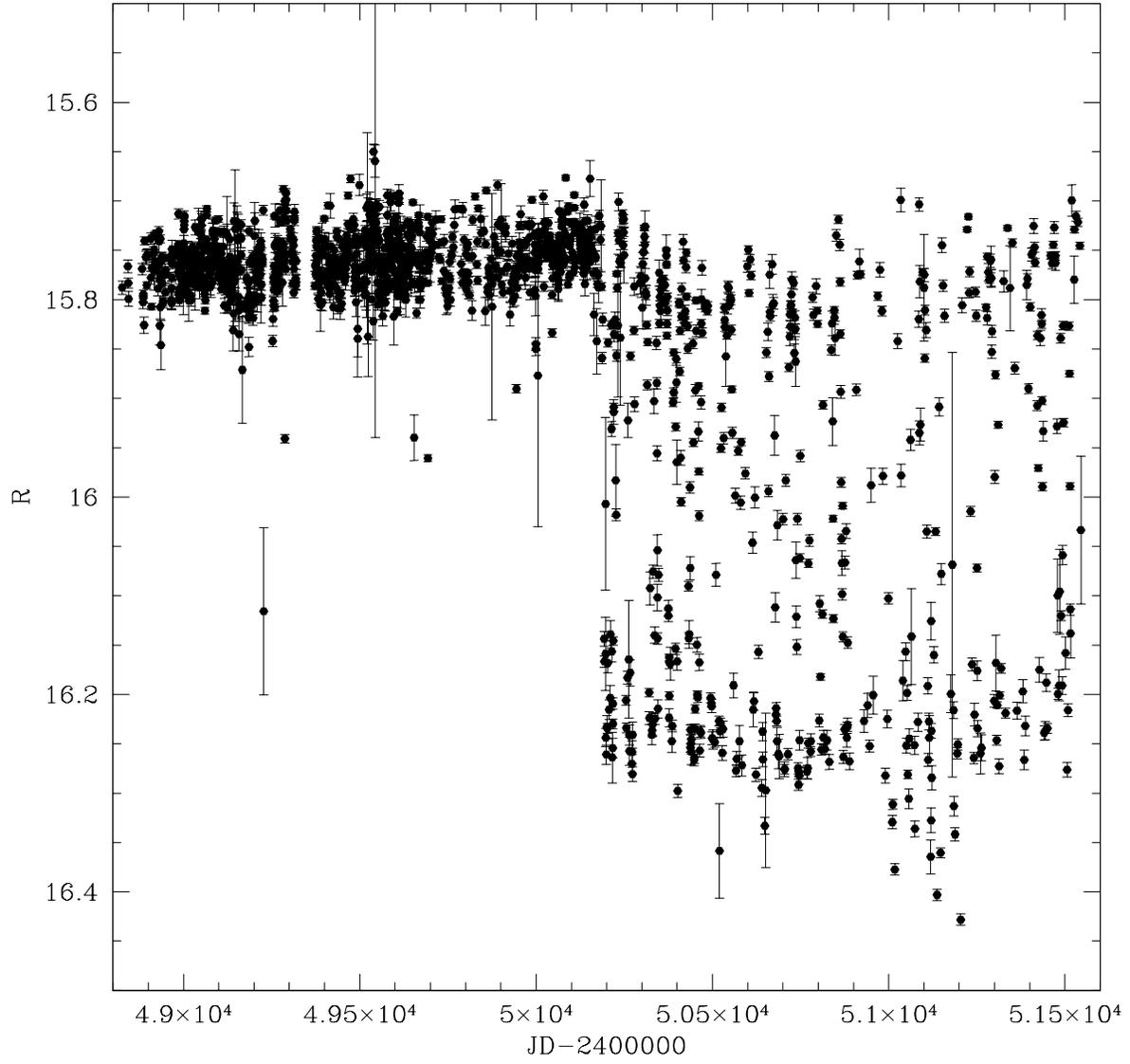}
\caption{$R$~light curve of MACHO~78.5979.72. Times are given in 
JD-2400000.}
\label{fig:fig1}
\normalsize
\end{center}
\end{figure}

Figure \ref{fig:fig1} shows the variable's MACHO $R$ light curve for 
a continuous time span of about 7.4 years (save for a 60-day interval 
between November 1993 and January 1994, due to telescope problems) 
derived from the MACHO database.
The MACHO~instrumental photometry has been calibrated to the standard 
Kron-Cousins system (K. Cook, private communication).
\par
The light curve exhibits several features, which, so far, are unique 
among LMC~variables, particularly among its "bumper" variables, and 
galactic variables as well.
\par
\begin{figure}
\begin{center}
\footnotesize
\plotone{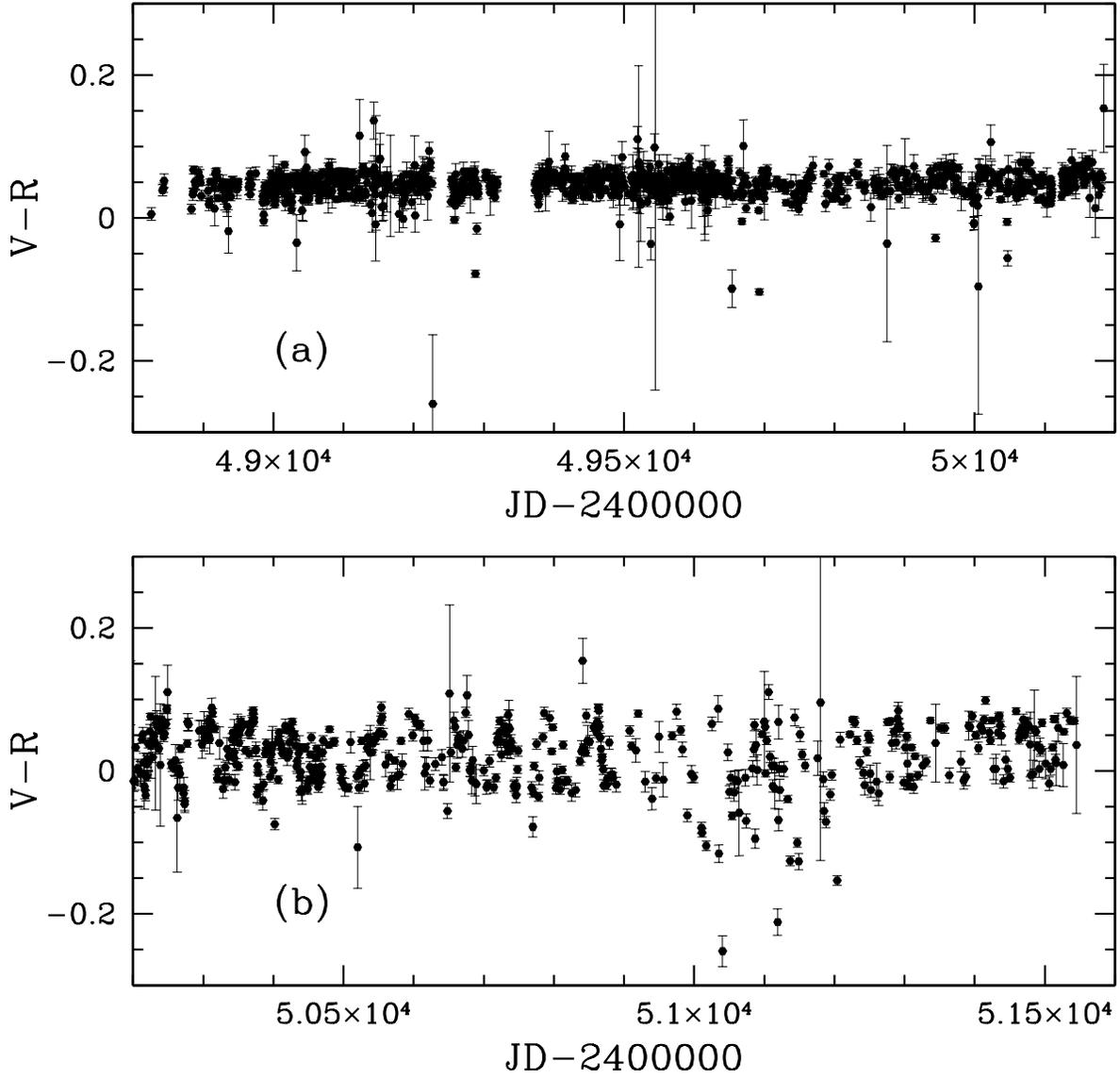}
\caption{$\vr$~light curves for (a) the quasi-flat segment and (b) 
the periodic segment separately. 
Times are given in JD-2400000.}
\label{fig:fig2}
\normalsize
\end{center}
\end{figure}
\begin{figure}
\begin{center}
\footnotesize
\plotone{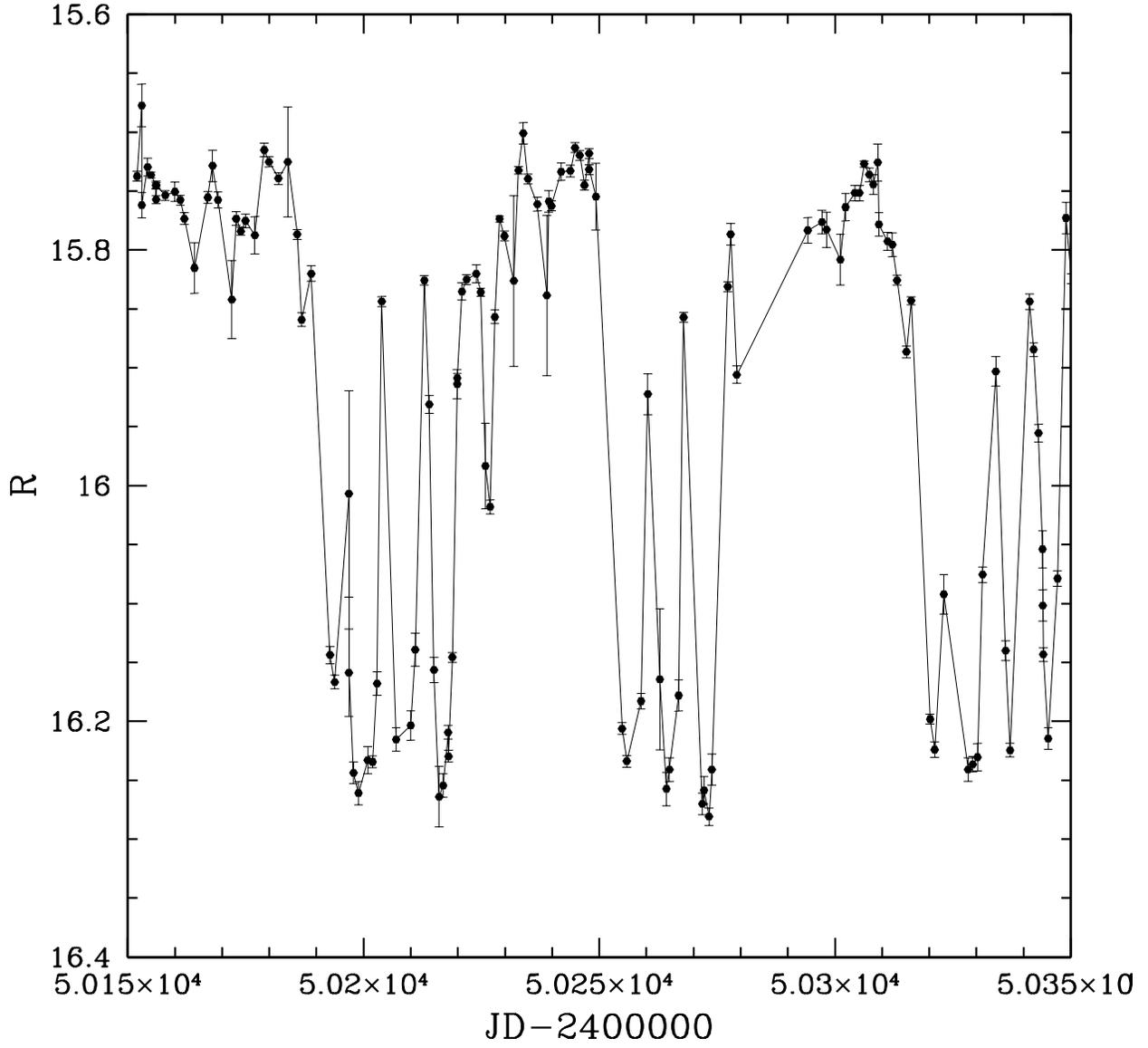}
\caption{$R$~light curve at a wider time resolution around the rapid
transitional onset at approximately JD2450180.}
\label{fig:fig3}
\normalsize
\end{center}
\end{figure}

We discuss several features of the light curves in turn.
\begin{enumerate}
\item
For approximately the first half of the time span there is no obvious 
periodic variability, while for the remaining half there is a periodic 
variability suggestive of an eclipsing phenomenon; we will refer to 
these as the quasi-flat segments and periodic segments respectively; 
Fig. \ref{fig:fig2} shows the $\vr$~light curves for each of these 
segments separately.
Errors in $\vr$ are the quadrature sum of errors in $V$ and $R$.
\par
Ironically, because planning strategies for some MACHO~fields changed 
about midway, the quasi-flat segment has about 950 observations per 
filter (about one per 1.5 nights) while for most of the periodic 
segment only about half as many were secured. The AGAPEROS data of 
1991 overlaps a small portion the quasi-flat MACHO~segment, and 
OGLE~II data covers only the periodic segment and none of the 
quasi-flat segment.
\par
The MACHO $V$ and $R$ total light curves, and the quasi-flat and 
periodic segments of the light curve, were separately subjected to a 
power spectrum analysis using the Lomb Periodogram technique, which is 
ideal for unevenly sampled data; notably, it is insensitive to gaps, 
periodic or random, in data \citep{press92}. Table \ref{tab:tab1} 
lists the most significant periods in the power spectrum (i.e. those with the
largest amplitudes) for the total light curve, the quasi-flat segment, 
and the periodic segment.
Uncertainties in these periods were estimated from the weighted dispersion of the derived frequencies in the power spectrum (Rice 2004, private communication). 
Two consistent significant periods of about 61 days and 8 days appeared in all of the data, except in the quasi-flat segment of the $V$~light curve where only the 8-day period was significant. As expected, the power spectra also show significant periods related to observational cycles of 0.5 and 1 day, and those related to the total observational time span (and some sub-multiples of it).
\par
The OGLE~II $I$,$V$, and $B$~band data for the variable, which 
cover much of the same time span as the periodic segment of the 
MACHO~data, were subjected to a Lomb Periodogram~analysis; the results 
are also listed in Table \ref{tab:tab1}. 
The only significant mean period in the $I$~band light curves 
(separately for DIA and DoPHOT photometry), is consistent with the 
61-day period found in the MACHO~data. For both the OGLE~$V$~and 
$B$~band light curves no significant periods were present.  
The object is not included in the OGLE~catalog of eclipsing binaries 
in the LMC~\citep{wyr03} presumably because of its peculiar light 
curve (i.e., it is not "clean") due to the superposition of the two 
periods.
The total temporal coverage between MACHO and OGLE data is 11.8 
years and the object is still monitored by the OGLE~team 
(Udalski 2004, private communication).
\item
The quasi-flat segment MACHO $R$ magnitude increases linearly at about 
0.01 mag yr$^{-1}$ while the $V$ magnitude increases linearly at about 
0.005 mag yr$^{-1}$, i.e., the system becomes secularly redder 
($\Delta(\vr)/\Delta t$= 0.013 mag yr$^{-1}$, cf. Fig. \ref{fig:fig2}) 
as it brightens in the quasi-flat segments.
\item
The envelopes of the maxima of the periodic segments are not constant, but become slightly fainter by $\sim 0.05$ mag at the onset of the variability in both 
$V$ and $R$; data sampling is unlikely to have caused this behavior in the MACHO data.
The envelope of the maxima of the OGLE~II $I$ band light curve is also not constant, showing the same peaks as the MACHO data.
The $\vr$ light curve of the periodic segment shows that the object is redder at maxima and bluer at minima; the scatter of points with
$\vr<0$ just past the middle of the data (JD 2451000), which begins 
about 850 days after the onset of complete eclipses and lasts for some 
200 days, are associated with deeper minima over this interval.
\item
The envelopes of the minima of the MACHO periodic segments also are 
not constant; data sampling is unlikely to have caused this 
behavior in the MACHO data; by comparison, the envelope of the minima 
of the OGLE~II $I$ band light curve is fairly flat.
\item
The onset of the periodic segment approximately midway in the total 
time span, described as a 0.6 mag drop in $R$ and a 0.45 mag drop in 
$V$ ($\Delta(\vr)\ge$0.15 mag, cf. Fig. \ref{fig:fig2}), simply just 
begins without antecedent indications of a gradual approach, as 
illustrated in Fig. \ref{fig:fig3} for the $R$ light curve, which is a
wider time resolution of this rapid transitional onset and, for 
clarity, the sequential data points are connected with straight lines.
This initial magnitude drop occurs over 10.0 days.
\item
The minima of the periodic segment contain three, and occasionally 
four, spikes of light of short duration (a few days) that appear 
periodic when viewed at the wider time resolution of Fig. 
\ref{fig:fig3}, and they appear to be symmetrically placed within the 
minima; the spikes generally do not become as bright as the quasi-flat 
segments or the maxima of the periodic segments, although there are 
occasional very bright spikes of light in the minima, such as that at 
JD 2451245. 
These light spikes are associated with the 8-day periodicity.
\item
The width of the observed minima of the MACHO light curve, estimated 
at half depth, widens as time progresses, from about 30-35 days at the 
beginning (cycles 1-5) to about 40-49 days around JD 2451080 
(cycle 15), some 2.5 years later. 
The overlapping OGLE-II data show the same trend, but after JD 2451140
(cycle 16) the width of minima begin to shorten again to about 28-38 
days.
\item
The most recent $I$ band OGLE-III data \citep{udalski03}
shows another secular change has occurred in the light curve 
of the object in the 1.2-year interval between end of the OGLE-II 
data (JD 2451690) and beginning of the OGLE-III data (JD 2452123). 
Coverage is sparser (between 1 and 18 nights per observation; average 
is about one observation per 5 nights) and the behavior is different 
from previous MACHO and OGLE-II data: maxima and minima appear 
shorter in duration, and without any evidence of short spikes of light 
(which are only a few days long) present in the periodic segment of 
the MACHO data. 
Lomb Periodogram analysis was applied to this data and the results are 
also listed in Table \ref{tab:tab1}: the only significant period was 
one of about 30.4 days, consistent with half the 61-day period within 
the uncertainties. 
It is unlikely that this short period is caused by sparser data 
coverage.
\end{enumerate}
\par
The object's non-variability in the Hodge-Wright study of 1967 
indicates its amplitude was $<$ 0.1 mag at this time, and so is 
consistent with the small amplitude during the quasi-flat segments of 
the MACHO~light curves.
This suggests either that the quasi-flat segment extends back this far 
back in time, or that it was imaged during a previous quasi-flat 
segment.
We have not found any images of the object in the interim between 1967 
and 1991 (AGAPEROS survey) to assess this latter possibility.  
The 2MASS images were taken at a maximum in the periodic segment.
\par
\begin{figure}
\begin{center}
\footnotesize
\plotone{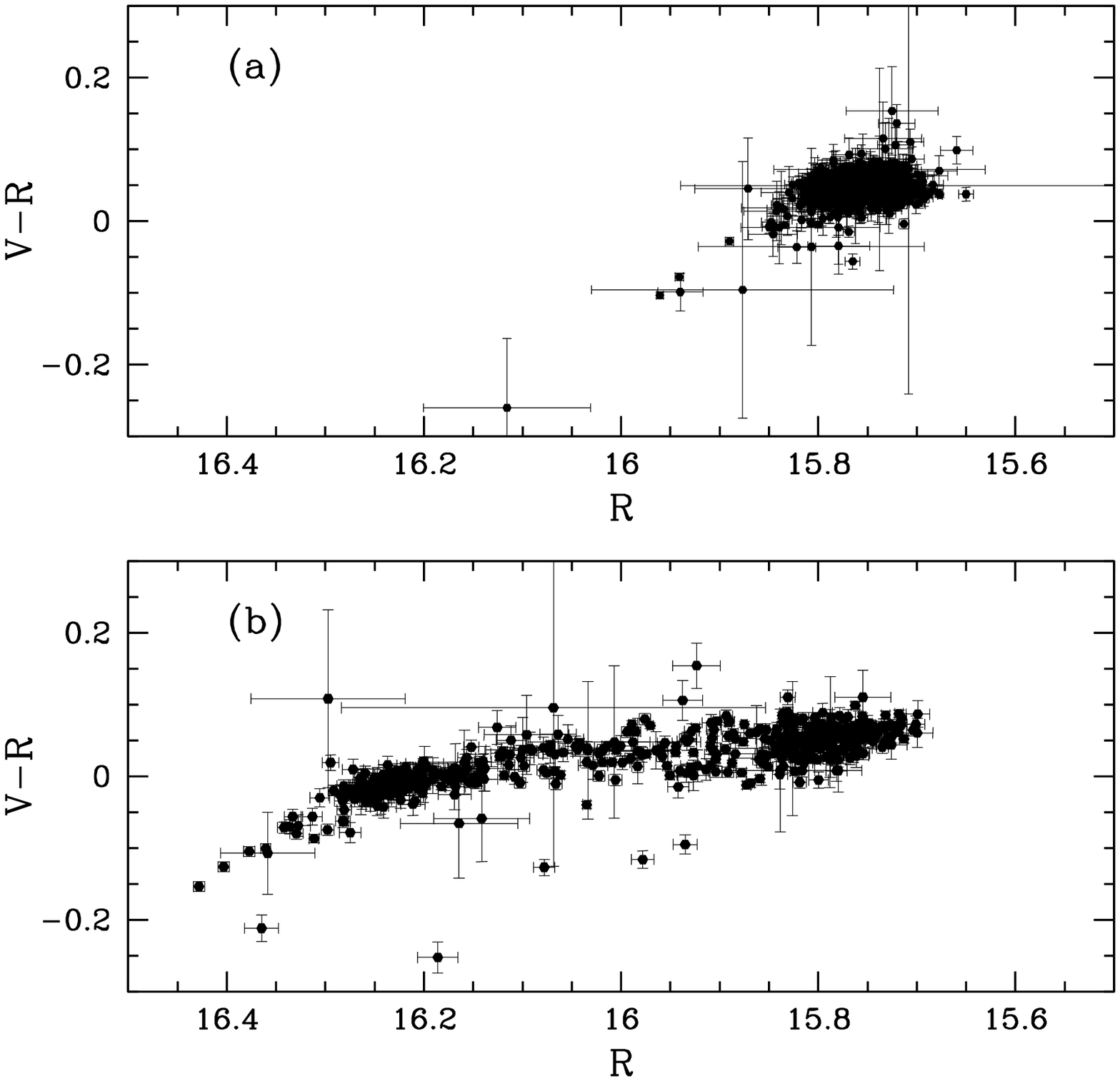}
\caption{MACHO~$\vr$~vs. $R$~magnitude for (a) the flat segment and
(b) the periodic segment. 
The system clearly becomes redder, on average, as it brightens.}
\label{fig:fig4}
\normalsize
\end{center}
\end{figure}

Figure \ref{fig:fig4} illustrates the MACHO~$\vr$ color vs. $R$ 
magnitude for the two light curve segments separately.
Despite the presence of outlying points, this plot clearly shows that 
maxima are 0.1 mag redder than minima in both segments of the light
curve.
\par
\begin{figure}
\begin{center}
\footnotesize
\plotone{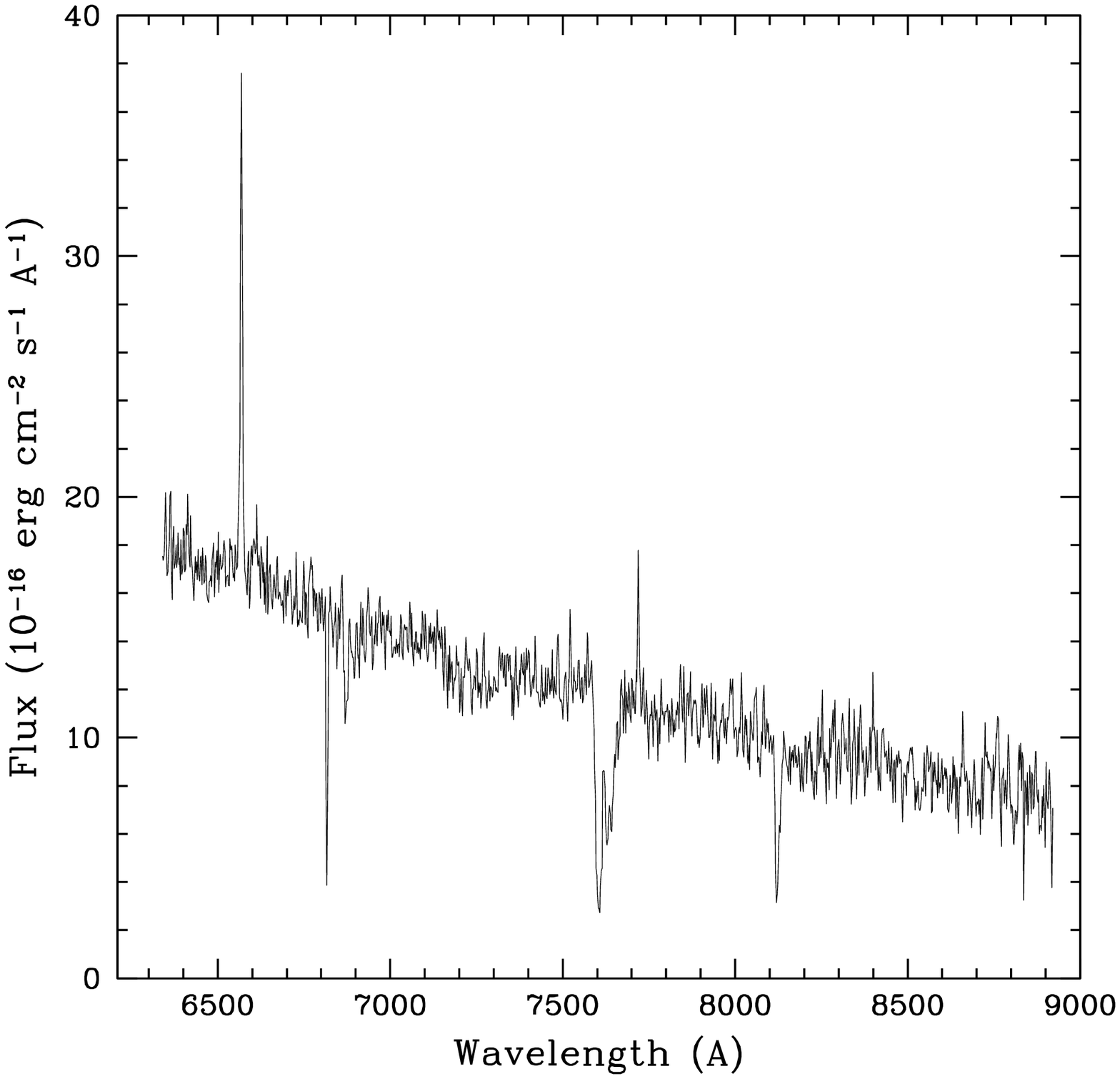}
\caption{A low resolution near infrared spectrum secured on 11/11/03. 
The only two significant stellar spectral lines are the strong H$_\alpha$ emission and Fe II $\lambda$7712 which are characteristic of a B2/3V/IVe spectrum. The ``absorption'' line near 6810 A is spurious, caused by a bad column in the CCD readout; other absorption lines are terrestrial.}
\label{fig:fig5}
\normalsize
\end{center}
\end{figure}

Five low resolution near infrared\footnote{Some authors prefer the term ``far red" for wavelengths between 0.7$\mu$m and 1.0$\mu$m.} spectra of the variable were secured with the Cassegrain spectrograph of the CTIO 1.5-meter telescope on two nights: three on Nov. 7, 2003 (JD 2452950.80) which are very noisy, and two on Nov. 11, 2003 (JD 2452954.76) which are much less noisy.
Figure \ref{fig:fig5} shows the reduced spectrum for the second night.
The spectra appear to be those of a B2/3V/IVe star, with H$\alpha$ emission of equivalent width of about 10 A at both times; they are comparable to those in the low resolution near infrared spectra in \citet{mennickent97}, and with the higher resolution spectra of \citet{andrillat88}.
\par
Consistent with expected behavior of this sample of blue variables, the H$\alpha$ emission lines appear when the variable ascended toward a maximum on the first night and at that maximum on the second of the 
recent OGLE-III light curves. 
The radial velocity of the H$\alpha$ line is $\sim 280\pm 40$ km s
$^{-1}$ on the first night and $\sim 240\pm 40$ km s$^{-1}$ on the 
second, and that of the Fe II emission line $\lambda 7712$, the only 
other obvious emission, is 290$\pm$20 km s$^{-1}$ on both 
nights.\footnote{The presence of Fe II $\lambda 7712$ without O I $\lambda7774$ also in emission, as in our variable, is rare in some Be star samples \citep{andrillat88}, but there are exceptions, eg. o Cas (HD 4180), a B5IVe star \citep{andrillat88}, and $\beta^1$ Mon \citep{polidan76}, a B3Ve star.}
All radial velocities are consistent with membership in the LMC.
Uncertainties in the radial velocities on the two nights are too large to draw any firm conclusions regarding variations between them.
There are no clearly detectable stellar absorption lines.
\par
Since the variable's absolute magnitude, color, and spectra are all
consistent with an early Be star, interstellar reddening along its line
of sight appears to be low.
\section{Phase Diagrams}
\label{sec:phasediag}
Phase diagrams of the MACHO data were constructed with period of 61.295 days for \textit{both} the $V$ and $R$ quasi-flat segments (even though this period was found only for the $R$ data), and with a period of 61.462 days for the periodic segments in both filters.
The $R$ phased light curve is shown in Fig. \ref{fig:fig6} for both segments.
The phase diagrams exhibit several distinctive features:
\begin{figure}
\begin{center}
\footnotesize
\plotone{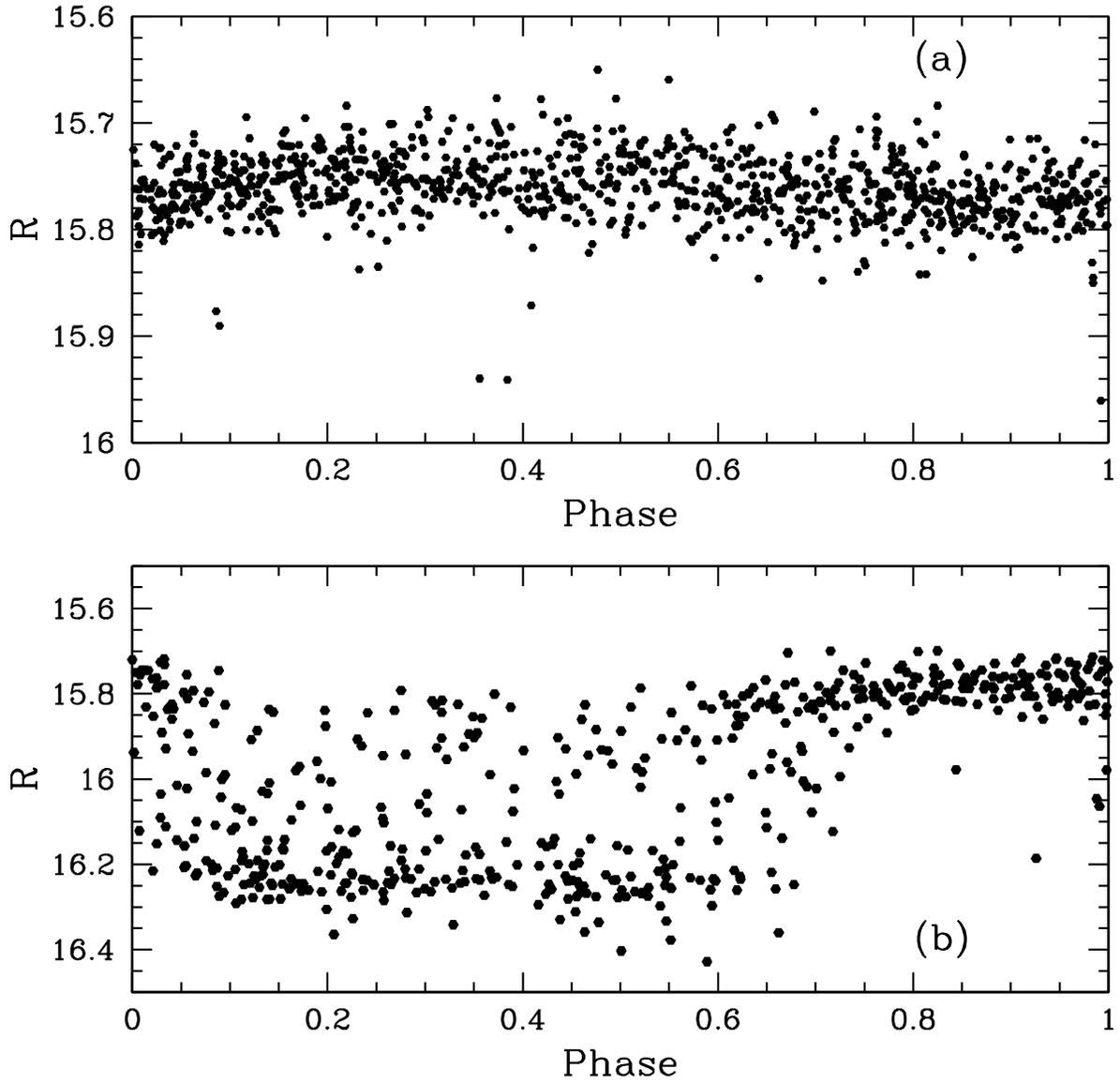}
\caption{(a) 61-day phase diagram of the quasi-flat segment of the 
$R$~light curve; $P=61.295$~days and $T_{0}$=2450000~days. 
(b) 61-day phase diagram of the periodic segment of the $R$~light 
curve; $P=61.462$~days and $T_{0}$=2450000~days.}
\label{fig:fig6}
\normalsize
\end{center}
\end{figure}
\begin{enumerate}
\item
In the periodic segments the minima occupy $>50$\% of the phase diagrams in both $V$ and $R$, while the maxima occupy only about 30\%; the shoulders of the minima occupy the remaining fraction and are not well defined.
Such a large and ill-determined fraction can be caused by a variable widening of the minimum and consequent narrowing of the maximum, as well as changing location of the center of the minima, but no periodicities are associated with these possibilities.
The phase diagram for $\vr$ in the quasi-flat segment appears constant in color over the entire time span, while the minima of the periodic segment are bluer in the mean than the maxima by $\sim 0.15$ mag.
\item
In the periodic segments, the approximately flat-bottomed minima are 
randomly occupied with points partly arising from the 
incommensurability of the 8-day period with the 61-day period. 
\item
In the quasi-flat $R$ segment, the phase diagram is expected to be 
noisy due to the secular brightening noted above, and from  the 
incommensurability of the 8-day period with the 61-day period. 
The phase diagram appears to be effectively U-shaped, dropping by 
$\sim 0.05$ mag and is approximately 0.5 out of phase with respect to the minimum of the periodic segment.
The phase diagram in $V$ is flat with a dispersion of about 0.05 mag.
\end{enumerate}
\begin{figure}
\begin{center}
\footnotesize
\plotone{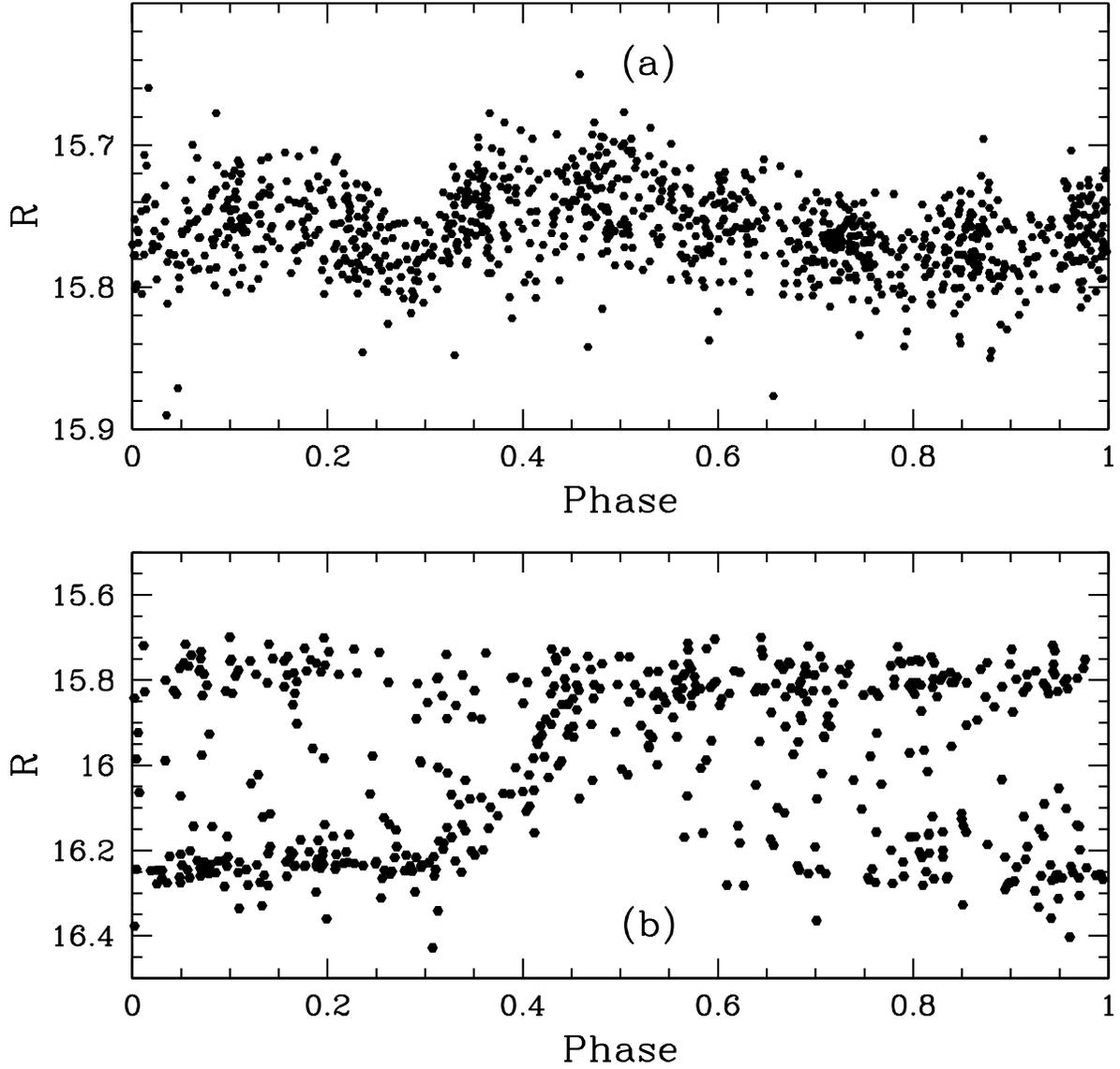}
\caption{(a) 8-day phase diagram of the quasi-flat segment of the 
$R$~light curve; $P=8.014$~days and $T_{0}$=2450000~days. 
(b) 8-day phase diagram of the periodic segment of the $R$~light 
curve; $P=8.010$~days and $T_{0}$=2450000~days.}
\label{fig:fig7}
\normalsize
\end{center}
\end{figure}
\begin{figure}
\begin{center}
\footnotesize
\plotone{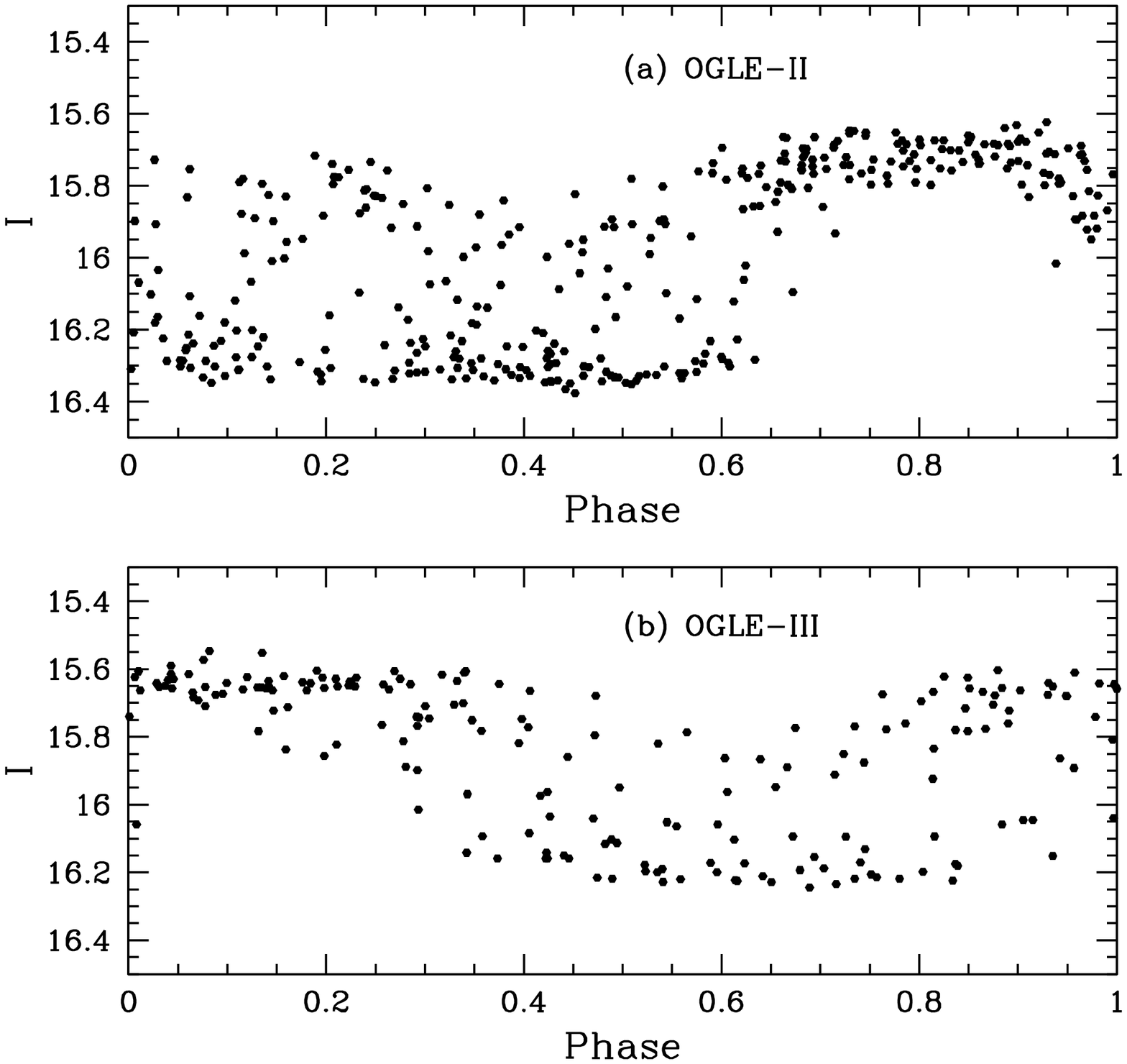}
\caption{(a) 61-day phase diagram of the periodic segment of the 
OGLE-II $I$~light curve; $P=61.642$~days and $T_{0}$=2450000~days. 
(b) 30-day phase diagram of the periodic segment of the OGLE-III $I$~light curve; $P=30.437$~days and $T_{0}$=2450000~days.}
\label{fig:fig8}
\normalsize
\end{center}
\end{figure}
A second set of phase diagrams for each segment of the $R$~and 
$\vr$~MACHO data was constructed using the periods of 8.014 days in 
the quasi-flat segment and 8.010 days in the periodic segment; Fig. 
\ref{fig:fig7} shows the behavior of the $R$ data.
These phase diagrams exhibit a few distinctive features:
\begin{enumerate}
\setcounter{enumi}{3}
\item
In the quasi-flat segments the noisy spikes occupy about 30\% of the 
phase (about 2.5 days long) and are asymmetric, having a shorter rise 
to maximum followed by a more gradual fall to minimum, which seems to 
be flat within the noise; the incommensurability of the 8-day and 
61-day periods certainly contributes to the noisy appearance. 
The $\vr$~phase diagram shows this spike to be symmetric within the 
noise and bluer at maximum, contrary to the general behavior of the 
system to redden when brightening.
\item
In the periodic segments the spikes have a similar asymmetric shape 
discernibly occupying 30\% of the phase, despite noise that fills in 
the minimum arising from the incommensurability of the two periods. 
The $\vr$~phase diagram is noisy with a non-random pattern and shows no
clear correlation with either the $V$ or $R$ phase plots, save for a 
tightening of points after the spike.
\end{enumerate}
\par
The minima of both the quasi-flat and periodic segments thus behave 
similarly: the quasi-flat segment appears to be a lower amplitude 
version of the periodic segment. 
We note that they are reminiscent of the behavior of eclipsing 
binaries, where the former appears similar to partial eclipses, but 
with low amplitude spikes of light within it, while the latter appears 
similar to complete eclipses, with the same spikes of light, now of 
larger amplitude, within it.
\par
The next periodogram analysis was applied only to the \textit{maxima} 
of the periodic segments; for $V\leq 15.865$ and $R\leq 15.760$ 
(containing 165 data points) \textit{no} significant period was 
present, which seems surprising. 
The absence of the 8-day period at the maxima thus means it is present 
\textit{only} within the minima. This fact infers that a pulsating star
cannot cause the 8-day periodicity, for its presence would be observed 
at maxima when both stars are visible. 
It is not a Cepheid because the system is both too faint and too blue.
The period-luminosity relations derived for LMC Cepheids 
\citep{madore91} predicts $V=-4.43$ and $\vr=0.37$ for an 8-day
period; including the 0.25 mag scatter of the data in  quadrature does 
not change this conclusion.
The 0.2-0.4 mag amplitude of the imputed pulsation is larger than 
expected for typical main sequence stars, but is not unexpected for 
periodic photometric variations observed for Be stars 
\citep{kitchin82}. 
\par
This result suggests we apply a periodogram analysis to the 
\textit{minima} of the periodic segments as well to see if a similar 
finding obtains. 
We applied the analysis for $V\geq$ 16.144 and $R\geq$ 16.142 
(containing 195 data points); again, \textit{no} significant period was
present. 
Coupled with the result found in the previous paragraph, this result 
suggests that the strength of the 61-day period is produced 
\textit{not} from the location of the minima or maxima but from the 
spikes with an 8-day period, i.e., the first, second and third spikes 
in the first minima are separated by about 61 days from the first,
second and third spikes, respectively, in the second minimum, which are
in turn separated by about 61-days the first, second and third spikes,
respectively, in the third minimum, etc.  
Since the minima change width, the contribution of both ingress and egress 
times to this periodicity cannot be discerned.
The approximate orbital synchronization of light spikes in the minima places limits their possible origin, and clearly eliminates the possibility of a pulsating star.
\par
Since the same two periods are present in the quasi-flat segment, we infer that this segment is, again, just a lower amplitude version of the phenomenon observed in the periodic segment: the 8-day period arises from lower amplitude ``spikes'' of light within U-shaped partial minima, and the strength of the 61-day period arises from the light spikes via the same approximate orbital synchronization.
\par
A third set of phase diagrams were constructed for the OGLE-II and OGLE-III $I_{DIA}$ band (hereafter referred to as $I$) and are shown 
in Fig. \ref{fig:fig8} (a) and (b) respectively.
Both look much the same as Fig. \ref{fig:fig6} (b) for the MACHO periodic segment with a noisy minimum that extends $>50$\% of phase, except that the latter uses only a 30.437 day period.
The phase of the latter appears to be $\sim 0.2$ out of phase with the
former.
\section{Ansatz}
The MACHO data for this system exhibits two significant periodicities, 
a long one of some 61 days and a shorter one of about 8 days. 
Orbital motion is one natural possibility responsible for the 61-day 
periodic variability of this object and is the basis for our models, 
which are at present largely phenomenological.
A time scale of 4 years or longer is needed to account for the rapid 
transition from a quasi-flat to a periodic light curve.
\par
One main ingredient in our models is a disk surrounding a Be star which
produces variations in its light and color.
Since Be stars do not follow the standard reddening line, the red colors of the Be stars are not due to absorption by circumstellar dust typical of interstellar dust \citep{keller98}, and so is attributed to emission by ionized hydrogen gas in the disk.
The light curve of the object indicates maxima are redder, contributed 
by H$\alpha$ emission, while minima are bluer because of its absence, 
so we conclude that disk emission contributes to the total luminosity 
of the system. 
Disks have been detected interferometrically around seven Be stars; 
they range in size between 5-10 stellar radii, are relatively thin, 
are infrared emitters, and their inclinations agree with the projected 
rotational velocities of their central stars, i.e., they are equatorial
structures \citep{quirrenbach97}. 
One kinematic feature of disks around Be stars is the observational 
deduction that they are in near-Keplerian rotation about them \citep{cassinelli02};
\citet{lee91} show that matter from the star's equatorial surface can drift outward subsonically, producing a thin Keplerian disk; it is called the viscous decretion disk model.
Disks are also likely to be responsible for modulating the observed luminosity of their parent star via variations in optical depth and irregularities in internal structure, such as spiral waves and warps \citep{balona00,baade00}.
Such disks are generally though to be excretion disks generated by mass loss by one component and less likely an accretion disk produced by mass loss from a companion.
\citet{cassinelli02} show that both the high density and angular momentum of disks around Be stars can arise from magnetic torquing.
The question as to whether Be stars are pre- or post-main sequence objects, or a mixture of both, particularly in the LMC, is unresolved \citep{keller98}.
\par
A second ingredient in our models is a mechanism to account for the rapid change from a quasi-flat to a periodic light curve.
Since the transition occurs over a single orbital period of about 10 days without any antecedent indications of a gradual approach, such as a progressive widening and deepening of eclipses, an impulsive change in orbital elements seems a more likely cause than a slow progressive one, such as might be caused by precession of a disk misaligned with a rapidly rotating, polar-flattened, central star.
We propose that an additional perturbing object in an elliptical orbit of long period of several years acts as an impulsive force that induces the rapid change from partial to complete eclipses at a close periastron passage.
\par
We first investigated the possibility that that system is an eclipsing 
binary. There are at least three galactic eclipsing binaries that have 
either ceased eclipsing (SS Lac, \citet{milone92,torres00}), changed 
eclipse patterns (AS Mus, \citet{soder74}), or have turned on and off 
sequentially (V907 Sco, \citet{sandberg99}). In none of the three cases
are observations continuous through the cessation or restart of 
eclipses as with our object. Further, in all three of these cases a 
third star in a wider, more eccentric, non-coplanar orbit about the 
inner close binary is either invoked or observed to be present to 
account for changes in its orbital elements and eclipse patterns.  
\par
Assume the system is a non-eclipsing binary with a disk surrounding the
primary star.
A dynamical change (i.e., a precession) of either the orbital elements of the secondary, or in the disk surrounding the primary, or more likely a coupling of both, causes the rapid transition to an eclipsing system.
The precession is caused by a third star with a much longer orbital period. The main difficulty with the eclipsing hypothesis is producing the $\sim 50$\% width of the minima.
The simplest application of two-body mechanics shows the only way to produce minima lasting 50\% of an orbital period is to require the primary companion's surrounding disk to be hard-edged, and to have a radius equal to the orbiting companion's circular orbit (i.e., at its periphery).
Eclipses are observed when the disk's line of nodes is orthogonal to the line of sight (i.e. they cannot be observed with a disk canted or edge-on).
There is the additional question of the stability of such an imputed disk with a secondary star near its periphery.
From the orbital period we attempted to model the variable's light curve by 
assuming a disk with appropriately placed annular gaps through which the orbiting secondary peers as it passes behind the disk responsible for the long minima.
This would account for the approximately symmetric placement of the 8-day spikes within the minima and the fact that they are approximately synchronized with the 61-day orbital period.
\par
This first model, however, has several failures that may be attributed 
to the ideal character of the disk we assumed. First, there is no 
obvious geometric way to produce the 8-day period during the quasi-flat
segments since the gaps are not visible. Second, the perturbation by 
the third star can produce dramatic changes in both the line of nodes 
and angle of periastron of the secondary companion's orbit, which 
induce large changes in the light curve that are not observed. These
additional variations led us to abandon further explorations of this 
model.
\par
The second model to account for the light curve of the variable is 
simpler and more successful.
It assumes, again, that a thin disk surrounds a B star, but with four obscuring sectors that orbit in unison with the 61-day period and are responsible for eclipses; equi-spaced gaps between the four obscuring sectors produce the 8-day period when the central star peers through them.
Such a model has two advantages: first, it can easily produce minima exceeding 50\% of the fraction of the orbital period by judicious choices of azimuthal sizes of the obscuring sectors, and second, variations in their location, width, height and optical depth can impart a ``noisy'' appearance to the minima.
A second mass with a longer orbital period is required to perturb the disk such that its line of nodes and angle of periastron initially produce partial eclipses and then quickly produces complete eclipses.
\par
Interestingly, there is a variable that exhibits behavior similar to our object.
The recently discovered pre-main sequence solar-like star KH 15D showed a single minimum some 3.5 mag deep lasting about 40\% of its phase in 2002, which has widened from about 30\% since its discovery in 1995; the object is also bluer at minimum than at maximum \citep{herbst02} like our variable.
A second interesting parallel of our system with this young object is the presence of extra light in the minima, characterized as a central light reversal that initially reached the same or higher magnitude as the maxima, but has declined in brightness in time as the minimum has widened.
\citet{herbst02} invoke the presence of a sharp-edged disk around a 
single star, or around the fainter companion of a binary, to account 
for both the single eclipse and the light reversal. 
High-resolution spectra by \citet{hamilton03} showed the system to 
be a weak-lined T-Tauri star surrounded by an accretion disk and 
possibly with a bipolar jet. \citet{johnson04} found the system was a 
single-lined spectroscopic binary with a period consistent with the 
photometric one.
One detailed model is presented by \citet{barge03}, who posit a single,
sharp-edged disk with a large-scale gaseous vortex (related to planet 
formation) that contains swarms of solid particles responsible for the 
deep eclipses. 
While this model can reproduce the mild central light reversal observed
more recently, it cannot reproduce their initial brightness 
(Barge 2003, private communication). 
A third interesting parallel with our object was secured from archival 
Harvard plate data from the early to mid $20^{th}$century. 
They indicate that the object showed no eclipses over this time span
\citep{winn2003}, although photometric data secured from Asiago 
Observatory plate material between 1967-1982 showed the system was 0.9 
mag brighter with shallower eclipses which were 
$\sim 180^{\arcdeg}$ out of phase with more recent ones 
\citep{johnson03}. \citet{winn04} and \citet{chiang04} modeled the 
system as a pre-main sequence binary eclipsed by a slowly moving opaque 
screen, suggested to be a precessing circumbinary disk or ring, which
is quite different from our model for the MACHO variable. 
Typical of some T-Tauri stars, a filamentary $H_{2}$ emission 
nebulosity appears to be associated with the object \citep{tokunaga04}.
\section{Model: Single Star Eclipsed by a Disk with Obscuring Sectors}
Our more successful model explaining the light curve of the variable 
is simpler than that assuming it is an eclipsing binary.
It posits a single B star surrounded by a thin gas disk with at least four 
obscuring sectors orbiting in unison at the 61-day period, which implies they 
are located at $\sim 0.5$-$0.7$ AU from the central star.
These sectors are geometrically portions of a ring; they are all equal
in angular size and the gaps between each have the same angular width.
The azimuthal location, width, height and optical depth of the sectors
govern the 8-day period and the appearance of the central star as it
sequentially peers between each of them, producing spikes within the
minima; they must be fairly sharp-edged in order to produce relatively
steep shape of the spikes which last 1-2 days.
These obscuring sectors could be comprised of larger particles that
dim but selectively scatter little of the central star's light.
Physically they could be dusty vortices such as that proposed for the 
deep minima of KH 15D.
The formation of dust-trapping vortices has been simulated in 
3-dimensional models of protoplanetary disks, and may be the sites of 
planetesimal formation \citep{johansen04}.
The remainder of the disk contains ionized hydrogen responsible for 
H$\alpha$ emission and reddening of the maxima.
\par
A companion star will truncate the disk around its primary (and vice versa) by gravitational interaction over time.
Simulations by \citet{artymowicz94} of the tidal/resonant truncation of circumstellar disks which are coplanar with the eccentric orbit of their parent stars indicate that truncation radii are smaller than for binaries in circular orbits because tidal forces are larger at periastron in an elliptical orbit than in a circular orbit of identical semi-major axis $a$.
For a reduced mass of 0.5 and eccentricity of 0.7 their simulations indicate truncation radii are $\lesssim 0.2a$, depending on disk viscosity.
Simulations of companions on orbits non-coplanar with a primary's disk truncate it similarly \citep{larwood96}.
For our model the reduced mass is 0.36 and $a=6.3$ AU, so the truncation radius is $\lesssim 1.2$ AU.
This suggests that while the obscuring sectors in the disk are located at a radius 
of $\sim 0.5$ AU, the periphery of the gaseous disk could extend to the truncation radius so that the obscuring sectors would not be at the disk's periphery; we have no observational constraints on the disk's extent.
Given a disk radius $R$ of about 1 AU seen nearly edge-on, the expected maximum velocity width can be estimated from $v=2\pi R/P\sim 120$ km s$^{-1}$, corresponding to a spectral line width of $\sim 8$ \AA, which is comparable to the equivalent width of the observed $H\alpha$ line.
The general noisiness of the observations, which we attribute to turbulence in the disk, is not modeled.
\par
While there is no obvious cause for the asymmetric location of the four obscuring sectors on half of the disk, we state simply that this circumstance is needed for our kinematic model over the span of the MACHO and early OGLE data.
Since more recent OGLE data indicate the system still shows eclipses, but with a period of only 30 days, then if obscuring sectors do produce eclipses of the central star, their location and/or characteristics have changed.
Disks around nearby Herbig Ae/Be stars imaged in near- and mid-infrared and submillimeter bands (Vega, $\beta$ Pictoris, Formalhaut, HR 4796A, and HD 141569) show asymmetric appearances, and some variability, attributed to clumps, possibly caused by perturbing effects of planets
\citep[cf.]{wyatt99,ozernoy00,mouillet01,holland03}.
Several theoretical studies  \citep[][and references therein]{barge95,johansen04}
find that large particles within vortices may lead to clumping and planetesimal formation.
\par
These systems with hotter central stars have some closer similarities to our variable than KH 15D.
The extended/asymmetric feature in the disk of HD 141596, a Herbig B9.5Ve star with molecular CO and H$_{3}^{+}$ emission, has been modeled as a particle-accreting anticyclonic vortex, which could be the progenitor of a gas giant planet.
\citet{delafuente03} argue that such vortices are more effective at capturing solid material than equivalent structures around solar like stars, such as KH 15D, making them components of protoplanetary disks.
The star appears to be a member of a triple system with M2 and M4 companions \citep{weinberger00}, although they would have little or no effect on the imputed vortex in the primary star's disk \citep{delafuente03}; however \citet{augereau04} show that a companion in a highly eccentric orbit can account for the disk's structure.
\par
We establish a coordinate system centered on the B star.
The x-axis lies along the line of sight with the positive direction away from earth, the y-axis is in the plane of the sky normal to the x-axis, and the z-axis is normal the xy-plane.
The azimuthal variable $\phi$ is in the disk plane, measured from the +x-axis. The disk surrounding the B star is tilted with respect to the xy-plane with the line of nodes adjusted to that the central star is initially unobscured by the disk.
We represent the optical depth of each of the four obscuring sectors as a sum of four compound exponential power laws representing the optical depth variation in the plane of the disk and perpendicular to it, plus a fifth term representing the optical depth variation of the entire interior of the disk including the gaps between the obscuring sectors, which depends only on the direction normal to the disk plane:
\begin{equation}
\label{eq:eq2}
\tau=
\sum_{i=1}^{4}
B_{i}
\exp{\left[-\left(\frac{s_{i}}{s_{a}}\right)^{m}\right]}
\exp{\left[-\left(\frac{\phi-\theta_{i}}{\lambda_{i}}\right)^{n}\right]}
+
B_{5}
\sum_{i=1}^{5}\exp{\left[-\left(\frac{s_{i}}{s_{a}}\right)^{m}\right]}
+
B_{6}+B_{7}\sin(\phi-\theta_5)
\end{equation}
where:
\begin{description}
\item[]
$B_{i}$ are coefficients determining the magnitude of the optical 
depth for each obscuring sector, and $B_{5}$ is a constant that is 
present in the light curve throughout the entire cycle, i.e., it 
describes the magnitude of the optical depth of the non-obscured gaseous part of the disk, $B_{6}$ and $B_{7}$ are coefficients determining the magnitude of the optical depth 180$^{\circ}$ out of phase with respect to the center of the dark sectors, and account for the variation in the quasi-flat segment;
\item[]
$s_{i}$ are distances normal to the disk plane, and $s_{5}$ is the normal distance from the central disk plane to an elemental area on the star; 
\item
$s_{a}$ is the axis scale height of the entire disk normal to the plane;
\item[]
$m$,$n$ are powers of exponential power laws, perpendicular to, and in 
the plane of the disk, respectively;
\item
$\phi-\theta_{i}$ are the angular separations between the line of sight and the center of each obscuring sector in the disk plane measured from the +x-axis;
\item[]
$\lambda_{i}$
are azimuthal angular scale factors of each obscuring sector along the 
disk plane centered at $\theta_{i}$;
\item[]
$\theta_{5}$ is the initial position of the maximum of the density enhancement in the gaseous part of the disk (corresponding to the quasi-flat segment).
\end{description}
\par
Models of disks around B stars that are contiguous with their equatorial
surface show that they become isothermal, with midplane temperatures between
$3000-10,000~\mathrm{K}$ within $\sim 50$ stellar radii \citep{millar99} precluding the formation of grains.
Some models of Be stars invoke a gap between the star's surface and the disk 
\citep{kitchin82,eisner04}, so we propose that the disk of our star does not extend to the star's surface, but has a gap of indeterminate width, since our formulation does not depend on any parameter related to its width.
Its appearance would be described as a wide ring such as modeled by
\citet{delafuente03}.
\par
The radial location of dust within a disk is governed by the standard dust
sublimation radius of a star, $R_d$, which is used to estimate the inner radius
in a disk where dust will survive at temperature $T_d$ for a star of radius $R_s$
and temperature $T_s$ \citep{monnier02}:
\begin{equation}
\label{eq:eq3}
R_{d}=\frac{1}{2}\sqrt{Q_R}\left(\frac{T_{s}}{T_{d}}\right)^{2}R_{s},
\end{equation}
where $Q_R$ is the ratio of the dust absorption efficiencies of the incident and
reemitted radiation field, i.e., $Q_R=Q_{abs}(T_s)/Q_{abs}(T_d)$.
Each $Q_{abs}$ is the dust absorption efficiency for a given grain size and
radiation field of color temperature $T$.
$Q_R$ varies between $1$ for large grains (as found for young stellar objects, 
YSOs) and $\sim 50$ for small grains \citep{monnier02}.
Assuming $T_d=1500~\mathrm{K}$ for silicate dust near sublimation,
$Q_R=1$, and $R_s$ and $T_s$ of the central B3 star given in \ref{tab:tab1}, 
yields $R_d\sim 1~\mathrm{AU}$, larger than estimated from the $61-\mathrm{day}$ period above.
\par
While Eq. (\ref{eq:eq3}) does not account for the presence of dust as close as 
$\sim0.5~\mathrm{AU}$ to the central hot star as needed for our modeling of the light curve, we remark that \citet{monnier02} noted a similar inconsistency for
YSOs.
They find that some YSOs with significant UV luminosity have inner disk radii,
measured via infrared interferometry, that are smaller than their computed
inner radii using Eq. (\ref{eq:eq3}), assuming silicate dust of 
$1~\mathrm{\mu m}$ size, or grey dust, at $1500~\mathrm{K}$.
They suggest a partial resolution of this discrepancy can be achieved if
low-density gas present in the gap between star and disk scatters UV radiation,
shielding dust in the disk from destruction and so permit it to survive closer
to the parent star.
Their admittedly simple evaluation of this mechanism shows it is most effective
for hotter stars but does not explain the inconsistency for some stars with luminosity $L>10^2~\mathrm{L}_\odot$ (see their Figure 3).
\citet{dullemond01} and \citet{dullemond03} have modeled an alternative structural feature of a disk: its inner rim can become inflated, so shadowing its outer parts and permitting dust to survive.
For the parameters of our star this self-shadowing model shows that dust will
not survive inside about $1~\mathrm{AU}$ so it does not overcome the inconsistency in our model.
Assuming no flaring geometry of the outer disk, this model has the advantage of
low infrared emission, as observed for our object.
\par
\begin{figure}
\begin{center}
\footnotesize
\epsscale{0.6}
\plotone{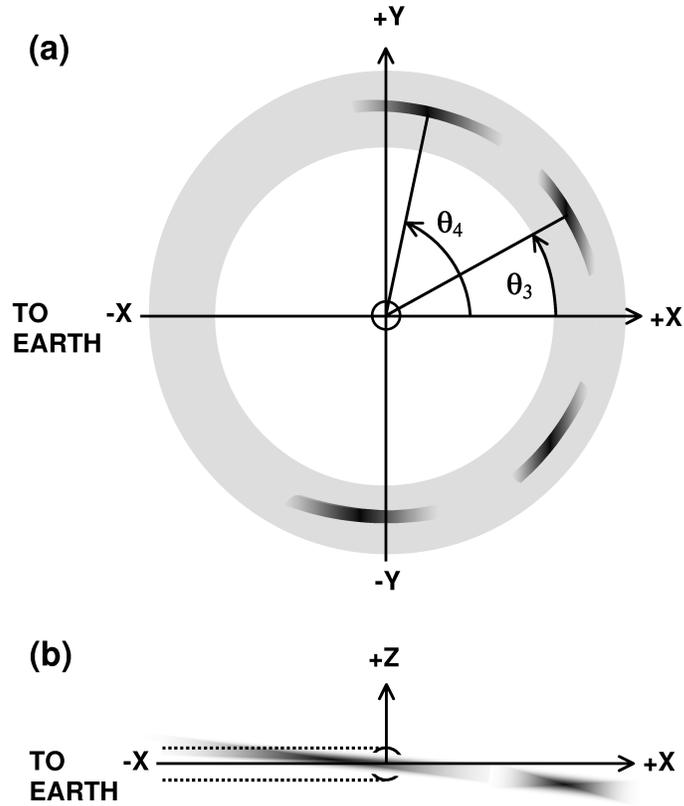}
\caption{Sketch of the model for the system. 
(a) Polar view of the disk, with obscuring sectors, surrounding a single star.
(b) Edge on view of the disk with obscuring sectors.}
\label{fig:fig9}
\normalsize
\end{center}
\end{figure}

Figure \ref{fig:fig9} illustrates some of the disk parameters for our model.
\par
At a given time, the radiant flux is determined by numerically integrating the flux emitted by a unit area in the line of sight over the star's surface, i.e.:
\begin{equation}
\label{eq:eq4}
F=\int_{-R_{s}}^{R_{s}}\int_{-R_{s}}^{R_{s}}F_{s}e^{-\tau}\mathrm{d} 
z\mathrm{d} y
\end{equation}
where $F_{s}$ is the radiant flux per unit area over the filter pass band, and $\mathrm{d} z\mathrm{d} y$ is a unit area on the star's surface;
the star is assumed to radiate as a black body.
Table \ref{tab:tab2} lists the parameters of the central star and its
surrounding disk that produced the best fit to the MACHO and OGLE-II photometric data.
\par
This model has the advantage that the length of minima can be any
value simply by adjusting the azimuthal widths of the obscuring sectors.
It can also more easily account for the noisiness of the minima: the
variation in the location, amplitude and width of the light spikes can
be adjusted by varying the width, height and optical depth parameters 
of the obscuring sectors.
The asymmetric shape of the light spikes in the minima (cf., 4 in 
section \ref{sec:phasediag}) could be caused by asymmetries in the 
optical depths of the leading and trailing edges of the adjacent 
obscuring sectors (although we do not model this feature).
We have assumed ideal obscuring sectors that do not vary spatially or temporally in azimuthal location or width, height, or optical depth.
This model also assumes that approximately half of the disk has no obscuring sectors but still can slightly dim the central star, and contributes red light from its H$\alpha$ emission.
We assume a smaller optical depth for the rest of the disk containing the obscuring sectors via the $B_{5}$ coefficient in Eq. (\ref{eq:eq2}).
\par
We model the cause of the rapid change from partial to complete eclipses as perturbations by a second object in a long period elliptical orbit that is highly inclined with respect to the disk.
Orbital elements that are much different from these, eg. short period, and/or circular, and/or coplanar with the disk, will not produce the rapid change required.
Perturbations in the Keplerian orbital elements of the obscuring sectors due to an impulsive acceleration as given by the Gaussian perturbation equations \citep{brouwer61,danby62} were integrated via a third-order Runge-Kutta routine.
\par
The circumstellar disk is inclined with respect to the xy-plane.
The line of nodes was adjusted at $120^{\circ}$ to the line of sight so that initially no light variations would occur.
The second, perturbing mass was assumed to have a very long period and a large orbital eccentricity.
The major effect of this arrangement was a rapid rotation of the disk's line of nodes to a direction close to the line of sight, which enabled the sectors to obscure the B star and initiate the light variations of about 0.5 mag.
The angular size of each sector, as seen by the B star, and their azimuthal locations were chosen to fit the width of first deep minimum at JD 50200 days (the Julian Date is given with 2,400,000 subtracted out in all subsequent discussions and figures).
\par
\begin{figure}
\begin{center}
\footnotesize
\plotone{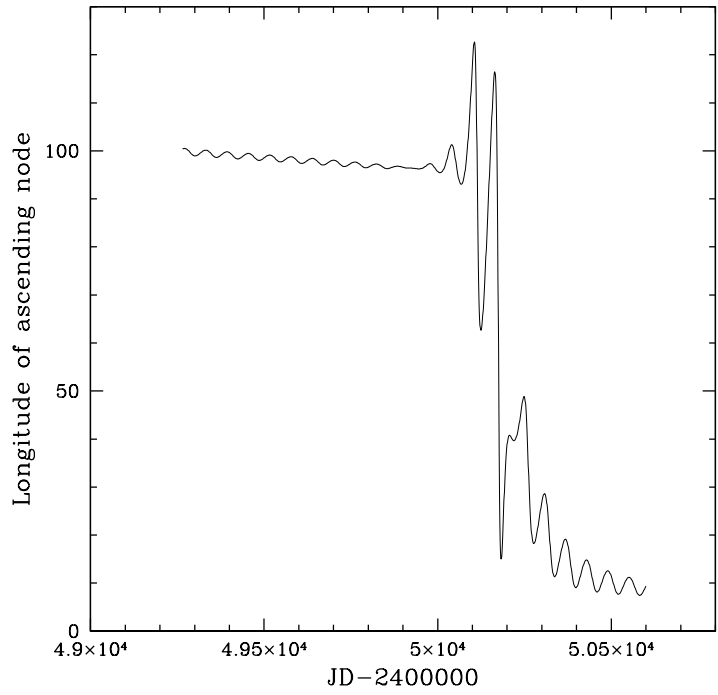}
\caption{Time evolution of the disk's line of nodes $\Omega$ in the 
model over the time span of the observations.}
\label{fig:fig10}
\normalsize
\end{center}
\end{figure}

Starting with an initial inclination, eccentricity and period, a large 
number of trials were carried out with different perturbing masses.
The initial argument of periastron, $\omega$, was always set to zero.
For each trial, the longitude of the ascending node, $\Omega$, was 
adjusted to have the deep minima begin to close to the observed times.
If the results were deemed not good enough, then the inclination was 
changed and the calculations repeated, for eccentricities ranging from 
0.5 to 0.8 and then for periods ranging between 1500 and 3000 days.
While a large number of trials were attempted, a definitive solution 
was not found.
The mass of the perturbing body was adjusted to produce a rate of
change of $\Omega$ such that minor light variations occurred for a
four-year span and then deeper minima ensued for the following
four-year span; Table \ref{tab:tab2} gives the orbital parameters for 
the perturbing mass that best fit the photometric data.
The best results were achieved with a mass of 3 M$_{\odot}$; a main sequence star of this mass contributes only a few percent to the total luminosity of the system.
Using a larger perturbing mass produced a slower turn-on of the 
periodic segment, which disagrees with the observations.
A minimum inclination of $-40^{\circ}$, i.e., retrograde motion with 
respect to the disk, is needed to produce a rapid change from partial
to complete eclipses caused by precession of the disk's line of nodes.
Figure \ref{fig:fig10} shows the time evolution of the disk's line of
nodes over this time period due to perturbation by this companion.
\par
The Gaussian perturbation equation for a test particle orbiting the central star of mass $M$, having semi-major axis $a$, eccentricity $e$, inclination $i$, distance $r$, and mean motion $n$, can be used to analytically estimate the precession rate of $\Omega$ induced by a perturbing mass $m$ \citep{danby62}:
\begin{equation}
\label{eq:eq5}
\frac{\mathrm{d}\Omega}{\mathrm{d}t}=
\frac{nar\sin(\omega+\nu)}{GM\sqrt{1-e^2}\sin i}N\approx
\frac{3nar^2}{\sqrt{1-e^2}R^3}\frac{m}{M}f,
\end{equation}
where $N$ is the perturbing acceleration due to $m$ normal to the disk's orbit, taken to be $3Gmrf/R^3$, and $R$ is its distance at periastron; $\nu$ is the test particle's true anomaly, and $f$ is a combination of sines and cosines of the angles between the orbits of the test particle and $m$, and is $\le 1$.
Eqn. (\ref{eq:eq5}) yields a maximum instantaneous nodal precession -rate of $\le 0.2$ deg day$^{-1}$, roughly consistent with our kinematic 
model results of $\sim 0.4$ deg day$^{-1}$, given the approximate values of the parameters used.
\par
We have treated the obscuring sectors as rigid portions of a ring, and the variation of other orbital elements give a variable rate of change of $\Omega$ as Fig. \ref{fig:fig10} shows; the \emph{mean} rate of change agrees with the formulation in \citet{larwood98}.
However, the gaseous disk precesses at a much slower rate than the dust particles; for comparable primary and secondary masses, \citet{papaloizou95} show $\Omega_{prec}\approx-(3/8)(n^2/\Omega)(3\cos^2i-1)$, which yields a precession $\sim 10^{-2}$ smaller for our model parameters.
If the total mass of dust particles is assumed negligible, our result based on the precession of the dust sectors may provide only an upper limit to the perturber's semi-major axis and/or a lower limit to its 
mass.
\par
Perturbation by the companion mass was assumed to effect all sectors
equally so the calculated widths of the deep minima do not change
during the entire timespan of the calculations.
The calculated amplitude of the light spikes was found to be sensitive
to the distances between the central plane of the disk and the center
of the B star as seen in projection along the line of sight.
Other disk orbital parameters also change ($e$ varies between 0.01 and
0.15, and $i$ varies between $-100^{\circ}$ and $-20^{\circ}$) with
smaller effects on the light curve
\par
\begin{figure}
\begin{center}
\footnotesize
\epsscale{0.55}
\plotone{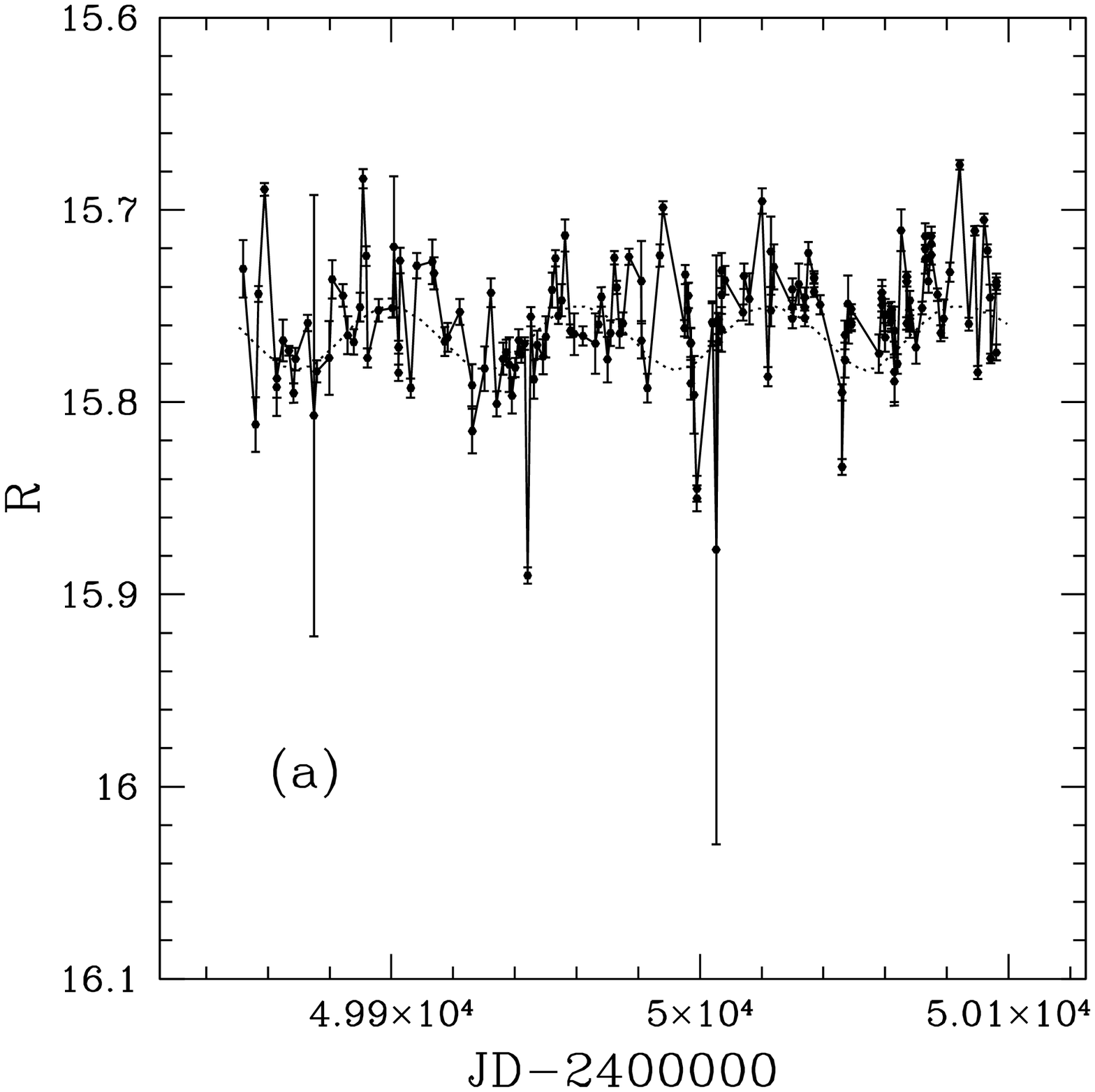}
\epsscale{0.55}
\plotone{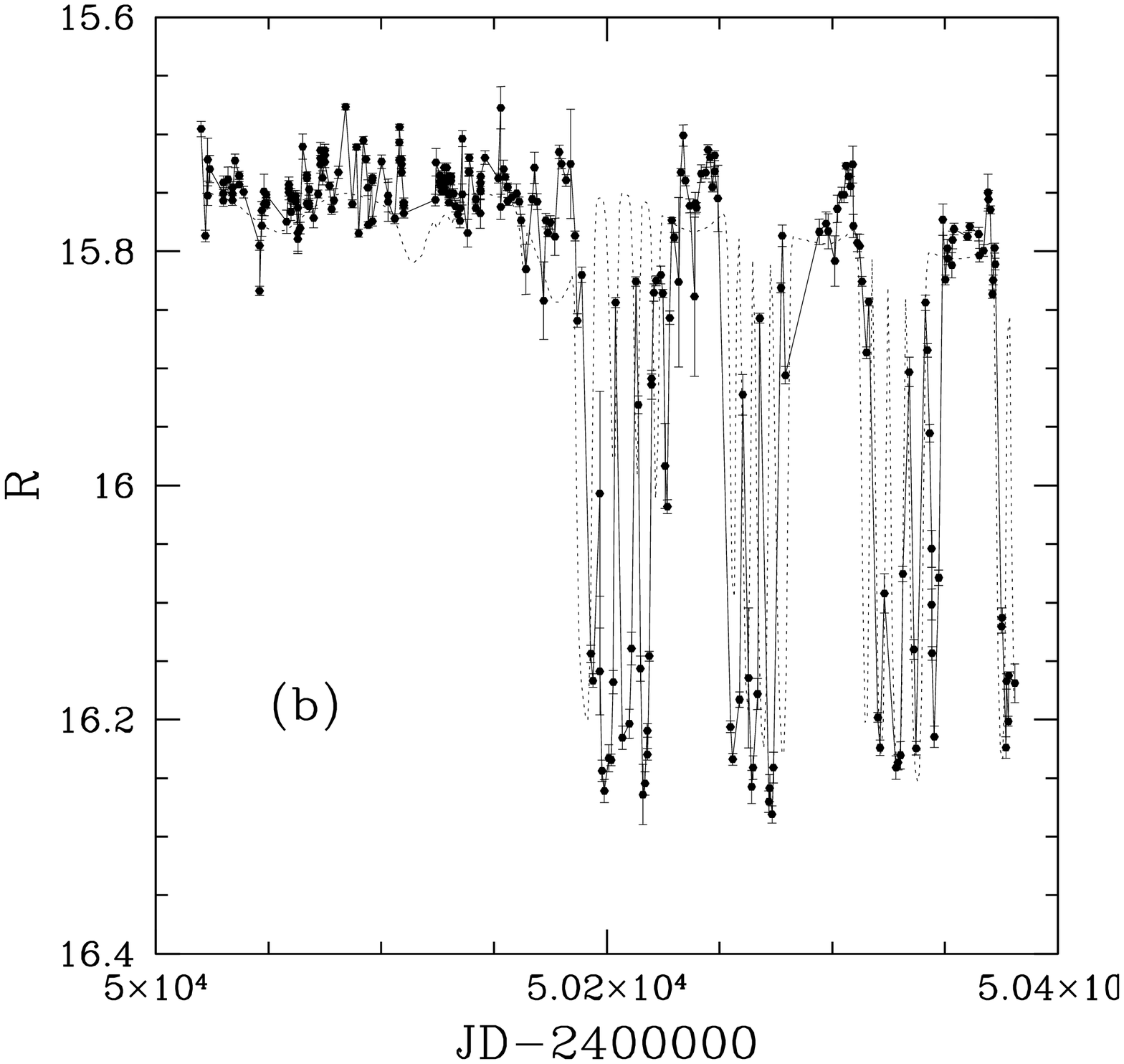}
\end{center}
\caption{(a) Overlay of a portion of the quasi-flat segment of the $R$ light curve
(solid line) with the predicted light curve (dashed line) of the model.
(b) Overlay of the observed $R$ light curve (solid line) 
prior to and around transition to the periodic segment with the 
predicted light curve (dashed line) of the model.
For both (a) and (b), $P=61.462$ days and $T_0=JD2450205$ days.}
\label{fig:fig11}
\normalsize
\end{figure}

Figure \ref{fig:fig11} illustrates the fit of our model with the observed $R$ light curve for a portion of the quasi-flat segment and around the transition time from a quasi-flat to a periodic light curve.
Inspection of the computed light curve vs. the data in Fig. \ref{fig:fig11} (b) informs us about the type of random variations in the obscuring sectors.
For example, we do not fit the timing of the light spikes in the minima very well; changes in location, width and optical depths of the obscuring sectors are likely responsible, but are not modeled.
The egress times from deep eclipse are fit reasonably well, indicating that the assumed period of 61.462 days is quite good and, in fact, contributes to the strength of this periodicity.
This behavior implies that the fourth obscuring sector is fairly fixed
in position with respect to the central star.
The same cannot be said of ingress times to deep eclipses: they vary
considerably, usually occurring earlier than observed, indicating that
the first sector has intrinsic random and secular changes we have not
modeled.
There are variations in actual times of ingress compared to the model:
starting at JD 50450 ingress occurs earlier, reaching a maximum 
separation at JD 51050, then approximately agreeing at JD 51175.
Relative to the model, the leading sector changes in size, location,
and optical depth, while the last sector remains relatively unchanging.
The changing width of the observed minima of the light curve noted
above suggests that the obscuring sectors are secularly changing: they
appear to have become larger and so cover a larger portion of the disk for about 15 cycles (about 920 days), but then may have decreased in size again as indicated by the recent OGLE~III data.
Our fit replicates the observational feature 3 of section \ref{sec:lc}: the envelope of the maxima becomes slightly fainter after the onset of the periodic segment; this is caused by the vertical structure of the disk modeled by the assumed exponential density falloff (of scale height $s_a$) as the precession of the line of nodes of the disk progresses.
\par
\begin{figure}
\begin{center}
\footnotesize
\plotone{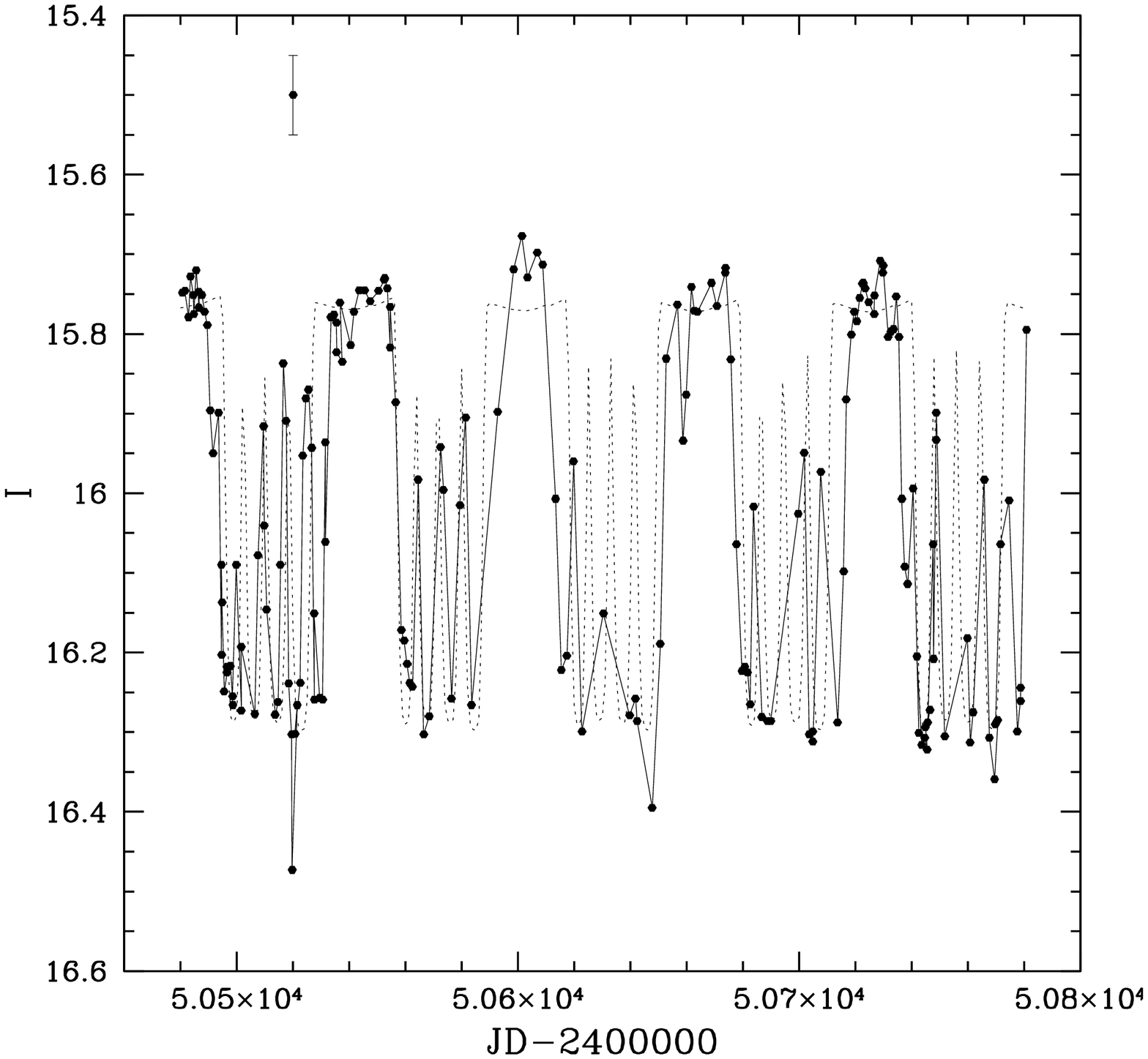}
\caption{Light curve of combined MACHO~and OGLE~II $I$ band data 
(solid line) for time a interval near the end of the periodic
segment overlaid with the predicted light curve (dashed line).
T$_{0}$ and $P$ are the same as those in Fig. \ref{fig:fig11}.
An average $1 \sigma$ error bar of the combined data is shown at the
upper left (MACHO errors are $\sim 0.03$ mag and OGLE errors are
$\sim 0.007$ mag).}
\label{fig:fig12}
\normalsize
\end{center}
\end{figure}

In order to fit our model to an observed light curve with a higher density of data points in the periodic segment, we combined the MACHO and OGLE~II data as follows. 
Using \citet{cousins80} $V$, $R$, $I$ photometry of the 23 galactic Be stars in his sample we derived the following color transformation:
\begin{equation}
\label{eq:eq6}
V-I=2.03(\vr)-0.02.
\end{equation}
We transformed the MACHO $V$ and $R$ magnitudes into $I$ magnitudes
with Eq. (\ref{eq:eq5}) and adjusted them to the zero point of the OGLE~II $I$ band data via the following transformation:
\begin{equation}
\label{eq:eq7}
I=V-2.03(\vr)+0.27.
\end{equation}
We then time ordered the derived $I$ band MACHO data and $I$ band 
OGLE~II data to produce a combined light curve.
Figure \ref{fig:fig12} compares the model with this combined light curve over a time interval in the periodic segment where the data points were most dense.
The widths of the minima and times of egress agree fairly well but the
8-day light spikes fit poorly.
The fit to the $I$ band data requires the $B_{1}$ through ${B}_{5}$ terms of Eq. (\ref{eq:eq2}) to be slightly larger than the fit to $R$ band data, suggesting that the dust particles comprising the obscuring sectors are large, consistent with the discussion 
implied by Eq. (\ref{eq:eq3}).
\par
We have only attempted to model the general features of the light curve
and the nominal behavior of the disk and the obscuring sectors.
We do not model random variations in the sectors, which could be caused
by turbulence, but they could be incorporated.
For example, the occasional fainter (and bluer) excursions near the 
middle of the periodic segment (cf., 3 in section \ref{sec:lc}, 
around JD 51000), could be caused by structure within the obscuring 
sectors.
The occasional very bright spikes of light (as at approximately JD 50820 and JD 51100) could occur if an occasional transparency in the disk occurs, permitting the central star to appear nearly unobscured.
\par
\begin{figure}
\begin{center}
\footnotesize
\plotone{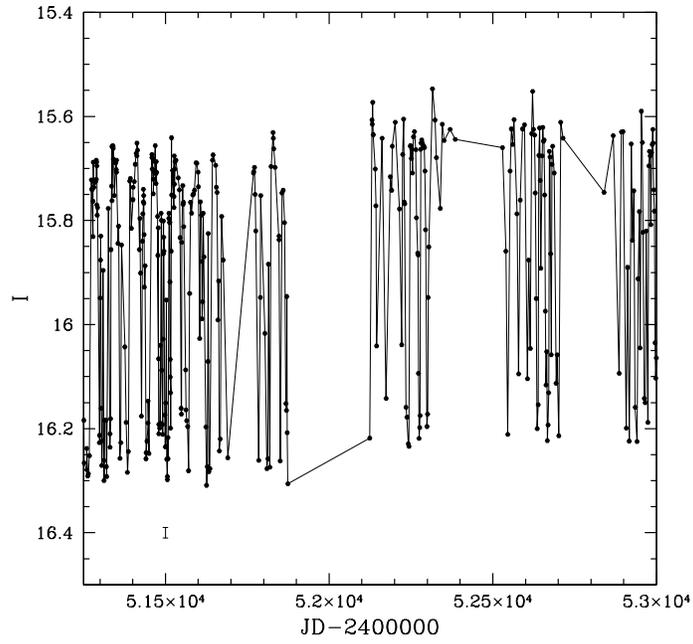}
\caption{Portions of the OGLE-II and III $I$ band light curve; a typical OGLE error bar is shown at lower left.
$T_{0}$ and $P$ are the same as those in Fig. \ref{fig:fig11}.}
\label{fig:fig13}
\normalsize
\end{center}
\end{figure}
\begin{figure}
\begin{center}
\footnotesize
\plotone{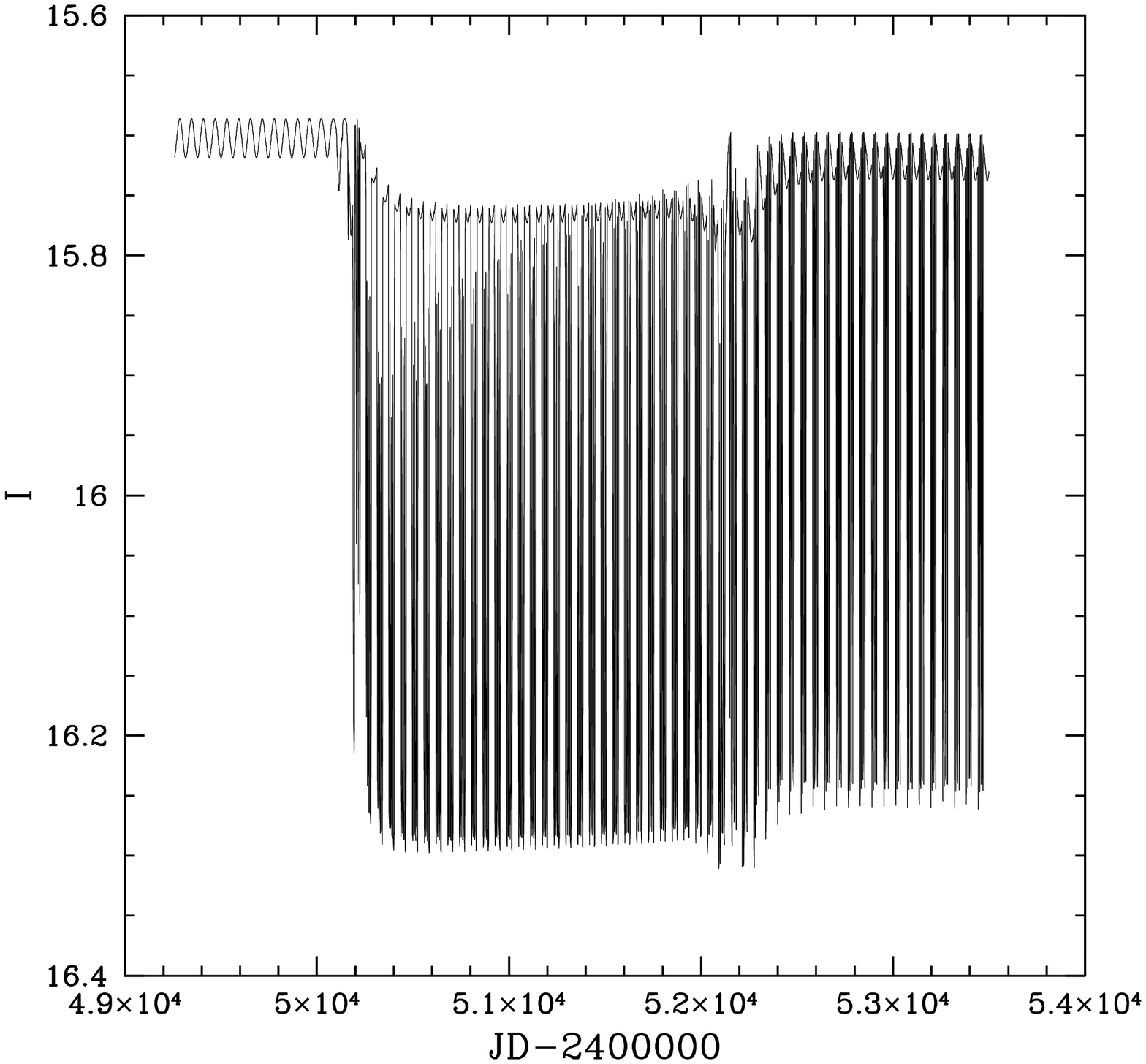}
\caption{Predicted time evolution of the $I$ light curve due to
the time evolution of the disk's line of nodes $\Omega$ over a time
span of $\sim 4300$ days.}
\label{fig:fig14}
\normalsize
\end{center}
\end{figure}

The recent OGLE-III (unpublished) data show only a 30 day periodicity is evident.
Although data are sparser, maxima are shorter and the three spikes of
light are not present within the minima, but are replaced by a single, wider maximum.
Figure \ref{fig:fig13} shows a portion of the OGLE-II and III data.
This new behavior, which occurred within $\sim 1.2$ years, could arise by assuming that the four obscuring sectors have coalesced into two larger, but unequal ones which have spread around the disk, such that gaps between each are now $\sim 180^{\circ}$ apart, permitting the central star to peer through each at approximately half the disk's assumed period of $\sim 61$ days.
Merging of vortices are seen in two dimensional models of compressible, viscous disks \citep{godon99,godon00}.
The newly coalesced obscuring sectors may also have moved radially inward slightly shortening their orbital period, perhaps due to turbulent motions.
Assuming OGLE~III $I$ magnitudes are directly comparable to OGLE~II $I$ magnitudes, then both maxima and minima of the former are about 0.1 mag brighter as Fig. \ref{fig:fig13} shows.
This behavior is predicted by our model light curve when the next periastron occurs, as Fig. \ref{fig:fig14} indicates.
But a caveat attends our assumption regarding direct comparison of OGLE~II and OGLE~III data: the latter are not as well calibrated as
the former and the mean magnitude could be the same, vitiating our
claim of a mean magnitude increase.
We have not modeled any of the detailed behavior seen in the OGLE-III data.
\par
We have inferred that much of the noisiness in the light curve could be related to the turbulent dissipation in the disk.
The time of a few years for the imputed disk of our object to secularly change its obscuring character is similar to that observed in KH 15D, and so may represent a relevant time scale for this process in stellar disks.
The decay of anti-cyclonic vortices in two dimensional models of viscous disks is 10-100 orbital periods \citep{godon99,godon00}.
Naturally, differences between our variable and KH 15D arise from
differences in the mass, luminosity and temperature of the central
star, as well as in parameters of their surrounding disks.
\section{Discussion}
Our model of light curve of the MACHO LMC blue variable FTS 78.5979.72 consists of a B star eclipsed by four obscuring sectors in a surrounding H$\alpha$ emitting ring-like disk.
A companion star in a retrograde long period eccentric orbit, inclined to the disk, is invoked to perturb the disk such that its line of nodes quickly changes the disk's orientation, and partial eclipses become complete ones.
We would hope that this Be variable is sufficiently interesting to impel disk theorists to employ hydrodynamical simulations, perhaps incorporating the main features of our simple kinematic model, which could represent more realistic physics of multiple vortices appearing asymmetrically on half a disk, as well as secular changes in a disk induced by the close passage of a companion star.
\par
The type of model we favor to explain the variability of this object, secular changes in a disk containing obscuring sectors which surrounds a parent B star, could be extended to explain some of the light curves of other variables in this sample and to those of Be stars in general.
While the line emission in Be star spectra is believed to arise from 
their circumstellar disks \citep{cassinelli02}, with the appearance of
emission lines at or near maxima \citep{dachs88,grebel97,keller98},
the additional feature of this class of LMC~variables noted by
\citet{keller02}, the redder maxima than minima, are additional inputs
for models of circumstellar disks.
\par
As to the predicted future of this system, two comments are relevant:
\begin{enumerate}
\item
The lifetime of disks with fully-formed planets is short because they
more easily dissipate them, and is estimated to be $\sim 10^{3}$
orbital periods; lifetimes are longer if planets are still forming 
within the imputed dusty vortices \citep{barge95,bryden00}.
No such planetary masses are included in our model.
\item 
The period of the companion in our model is unknown, but if it is shorter than the disk dissipation time scale, it could return to its
periastron, again exert perturbing impulses and further change the 
orientation of the disk.  We have run the eclipsing program for the 
model for longer times and find the following.
The disk's line of nodes could be moved some $90^{\circ}$ and the system would return to partial eclipses; if the obscuring sectors secularly change, the light spikes within the minima would change as well, as the recent OGLE~data suggests has occurred.
It is possible that the perturbing object is not bound to the system but has bypassed it at the time of the transition from the quasi-flat segment to the periodic, so the only future changes would be caused by the dissipation of the disk.
One prediction of our simple model with the parameters listed in Table \ref{tab:tab2} is that its velocity curve would be that of an
elliptical orbit, with an unprojected velocity at periastron of 
$\sim 30$ km s$^{-1}$ with respect to its centroid velocity.
\end{enumerate}
\par
Further spectroscopic data would clearly be of high value in refining the basic stellar and orbital parameters of this variable as well as
confirming or denying the fundamental model we have investigated.
Were spectral coverage sufficient to cover a few minima and maxima,
which are likely to be more similar to those in the recent OGLE~data
than those in the MACHO~data, further details of the disk's structure
and short-term secular evolution could be obtained.
\begin{deluxetable}{lcc}
\tabletypesize{\footnotesize}
\tablewidth{5.75in}
\tablecaption{Periodicities from Lomb Periodogram Analysis
\label{tab:tab1}}
\startdata
\tableline
\tableline
MACHO~datasets &
$R$-band P(days) &
$V$-band P(days) \\
\tableline
All Data (JD 2448825-2451544) & 
60.962$\pm$0.693 & 
61.514$\pm$0.584 \\
&
8.016$\pm$0.006 &
8.016$\pm$0.006 \\
Quasi-flat Segment (JD 2448825-2450185) & 
61.295$\pm$0.550 & 
\nodata \\
&
8.014$\pm$0.010 &
8.014$\pm$0.020 \\
Periodic Segment (JD 2450185-2451544) &
61.462$\pm$0.787 & 
61.462$\pm$0.796 \\
&
8.010$\pm$0.010 &
8.014$\pm$0.020 \\
\\
\tableline
\tableline
OGLE~datasets &
$I_{DIA}$-band P(days) &
$I_{DoPHOT}$-band P(days) \\
\tableline
OGLE~II (JD 2450457-2451690) &
61.642$\pm$0.778 &
61.572$\pm$0.680 \\
OGLE~III (JD 2452123-2453140)\tablenotemark{1} &
30.437$\pm$0.216 &
\nodata \\
\tableline
\enddata
\tablenotetext{1}{Kindly provided by Andrzej Udalski}
\end{deluxetable}
\begin{deluxetable}{ll}
\tabletypesize{\footnotesize}
\tablewidth{4in}
\tablecaption{Model Parameters\label{tab:tab2}}
\startdata
\tableline
\tableline
Central Star Parameter &
Value \\
\tableline
mass &
$5.6~\mathrm{M}_{\odot}$ \\
radius &
$4.33~\mathrm{R}_{\odot}$ \\
$T_{eff}$ &
$16000~\mathrm{K}$ \\
\tableline
\tableline
Circumstellar Disk Parameter&
Value \\
\tableline
Period &
$61.4618~\mathrm{days}$ \\
initial radius &
$116.3~\mathrm{R}_{\odot}$ \\
initial eccentricity &
$0.01$ \\
initial inclination &
$3^{\circ}$ \\
$\theta_{1-5}$ &
$0^{\circ}$, $45^{\circ}$, $90^{\circ}$, $135^{\circ}$, $67^{\circ}$ \\
$\lambda_{1-4}$ &
$21^{\circ}$ \\
$s_{a}$ &
$3~\mathrm{R}_{\odot}$ \\
$B_{1-4}$ & 
$0.50$ (for $I$ band) \\
& $0.42$ (for $R$ band) \\
& $0.40$ (for $V$ band) \\
$B_{5}$ &
$0.1$ (for all bands) \\
$B_{6}$ &
$0.05$ (for all bands) \\
$B_{7}$ &
$0.03$ (for all bands) \\
$m$, $n$ &
$6$ (for all bands) \\
\tableline
\tableline
Perturbing Object Parameter &
Value \\
\tableline
mass & 
$3~\mathrm{M}_{\odot}$ \\
Period &
$2000~\mathrm{days}$ \\
inclination &
$-40^{\circ}$ \\
ascending node &
$100^{\circ}$ \\
argument of periastron &
$-50^{\circ}$ \\
eccentricity &
$0.7$ \\
Time of periastron &
JD2450168.3 \\
\tableline
\enddata
\end{deluxetable}
\clearpage
\acknowledgments
We thank Andrew Drake for bringing this object to our attention, Kem Cook for the MACHO to Kron-Cousins magnitude transformations, Jeff Goldader for substantive comments and suggestions, Neil Reid for reducing the spectra, Andrzej Udalski for providing recent OGLE-III photometry, John Rice for discussions regarding frequency estimation and its errors, Pierre Barge for information on his disk model, and Megan Schwamb for help with some figures.
We also thank the referee for suggestions that clarified our kinematical
model.
This paper utilizes public domain data originally obtained by the 
MACHO Project, whose work was performed under the joint auspices of the U.S. Department of Energy, National Nuclear Security Administration by the
University of California, Lawrence Livermore National Laboratory under
contract No. W-7405-Eng-48, the National Science Foundation through
the Center for Particle Astrophysics of the University of California
under cooperative agreement AST-8809616, and the Mount Stromlo and
Siding Spring Observatory, part of the Australian National University.
This paper also utilizes public domain data obtained by the OGLE-II
project.


\begin{thebibliography}{99}

\bibitem[Andrillat, Jaschek \& Jaschek(1988)]{andrillat88}
	Andrillat, Y.,
	Jaschek, M., \& 
	Jaschek, C., 1988, 
	A\&AS, 72, 129

\bibitem[Artymowicz \& Lebow(1994)]{artymowicz94}
	Artymowicz, P., \&
	Lebow, S. H.,
	1994,
	\apj, 421, 651

\bibitem[Augereau \& Papaloizou(2004)]{augereau04}
	Augereau, J. C, \&
	Papaloizou, J. C. B.,
	2004,
	\aap, 414, 1153

\bibitem[Baade(2000)]{baade00}
	Baade, D.,
	2000,
	in ASP Conf. Ser. 214,
	The Be Phenomena in Early-Type Stars, 
	IAU Colloquium 175,
	eds. H.A. Smith, F.F. Henrichs, \& J. Fabregat,
	(San Francisco: ASP), 178

\bibitem[Balona(2000)]{balona00}
	Balona, L.A., 
	2000, 
	in ASP Conf. Ser. 214,
	The Be Phenomena in Early-Type Stars, 
	IAU Colloquium 175, 
	eds. H.A. Smith, F.F. Henrichs, \& J. Fabregat, 
	(San Francisco: ASP), 1

\bibitem[Barge \& Sommeria(1995)]{barge95}
	Barge, P., \&
	Sommeria, J.,
	1995,
	\aap, 295, L1

\bibitem[Barge \& Viton(2003)]{barge03}
	Barge, P., \& Viton, M. 2003, ApJ, 593, L117

\bibitem[Benedict et al.(2002)]{benedict02}
	Benedict, G.F., et al. 2002, AJ, 123, 473

\bibitem[Brouwer \& Clemence(1961)]{brouwer61}
	Brouwer, D., \& Clemence, G.M., 1961, 
	Methods of Celestial Mechanics, (New York: Academic Press)

\bibitem[Bryden et al.(2000)]{bryden00}
	Bryden, G.,
	Rozyczka, M.,
	Lin, D.N.C., \& 
	Bodenheimer, P., 2000, 
	ApJ, 540, 1091

\bibitem[Cassinelli et al.(2002)]{cassinelli02}
	Cassinelli, J.P.,
	Brown, J.C.,
	Maheswaran, M.,
	Miller, N.A., 
	Telfer, D.C.,
	2002, ApJ, 578, 951

\bibitem[Chiang \& Murray-Clay(2004)]{chiang04}
	Chiang, E., \& Murray-Clay, R.A., 2004, ApJ, 607, 913

\bibitem[Cook et al.(1995)]{cook95}
	Cook,~K. H. et al.
	1995,
	in ASP Conf. Ser. 83,
	Astrophysical Applications of Stellar Pulsation,
	IAU Colloquium 155,
	ed. R. S. Stobie \& P.A. Whitelock
	(San Francisco: ASP), 221

\bibitem[Cousins(1980)]{cousins80}
	Cousins, A. W. J., 1980, South African Astron. Obs. Circ., 1, 234

\bibitem[Dachs, Engels \& Kiehling (1988)]{dachs88}
	Dachs, J., Engels, D., \& Kiehling, R., 1988, A\&A, 194, 167

\bibitem[de la Fuente Marcos \& de la Fuente Marcos(2003)]{delafuente03}
	de la Fuente Marcos, C., \&
	de la Fuente Marcos, R.,
	2003,
	NewA, 8, 401

\bibitem[Danby(1962)]{danby62}
	Danby, H.M.A., 1962, An Introduction to Celestial Mechanics, 
	(New York: Macmillan)

\bibitem[Dullemond, Dominik, \& Natta(2001)]{dullemond01}
	Dullemond, C. P.,
	Dominik, C., \&
	Natta, A.,
	2001,
	\apj, 560, 957

\bibitem[Dullemond et al.(2003)]{dullemond03}
	Dullemond, C. P.,
	van den Ancker, M. E.,
	Acke, B., \&
	van Boekel, R.,
	2003,
	\apj, 594, L47

\bibitem[Eisner et al.(2004)]{eisner04}
	Eisner, J. A.,
	Lane, B. F.,
	Hillenbrand, L. A.,
	Akeson, R. L., \&
	Sargent, A. I.,
	2004,
	\apj, 613, 1049

\bibitem[Godon \& Livio(1999)]{godon99}
	Godon, P. \& Livio, M., 1999,
	\apj, 523, 350
	
\bibitem[Godon \& Livio(2000)]{godon00}
	Godon, P. \& Livio, M., 2000,
	\apj, 537, 396
	
\bibitem[Grebel(1997)]{grebel97}
	Grebel, E.K., 
	1997, 
	\aap, 317, 448

\bibitem[Hamilton et al.(2003)]{hamilton03}
	Hamilton, C.M.,
	Herbst, W.,
	Mundt, R.,
	Bailer-Jones, C.A., \&
	Johns-Krull, C.M.
	2003,
	\apj, 529, L45

\bibitem[Herbst et al.(2002)]{herbst02}
	Herbst, W., et al., 
	2002,
	\pasp, 114, 1167

\bibitem[Hodge \& Wright(1967)]{hodge67}
	Hodge, P.W. \& Wright, F. W. 1967, The Large Magellanic Cloud, 
	(Washington: Smithsonian Press)

\bibitem[Holland et al.(2003)]{holland03}
	Holland, W. S., et al,
	2003,
	\apj, 582, 1141

\bibitem[Johansen, Andersen \& Brandenburg(2004)]{johansen04} 
	Johansen, A.,
	Andersen, A.C., \&
	Brandenburg, A., 
	2004, 
	\aap, 417, 361

\bibitem[Johnson, Asher, \& Winn(2004)]{johnson03}
	Johnson, J.A., 
	Asher, J., \& 
	Winn, J.N.,
	2004,
	\aj, 127, 2344

\bibitem[Johnson et al.(2004)]{johnson04}
	Johnson, J.A., 
	Marcy, J.W., 
	Hamilton, C.M.,
	Herbst, W., \&
	Johns-Krull, C.M., 
	2004,
	\aj, 128, 1265 

\bibitem[Keller, Wood \& Bessel(1998)]{keller98}
	Keller, S.C.,
	Wood, P.R., \&
	Bessell, M.S., 1998,
	A\&AS, 134, 481

\bibitem[Keller et al.(2002)]{keller02}
	Keller, S.C.,
	Bessell, M.S.,
	Cook, K.H.,
	Geha, M., \& 
	Syphers, D., 2002, 
	AJ, 124, 2039

\bibitem[Kitchin(1982)]{kitchin82}
	Kitchin, C.R., 1982, Early Emission Line Stars, 
	(Bristol: Adam Hilger Ltd.)

\bibitem[Larwood et. al(1996)]{larwood96}
	Larwood, J. D.,
	Nelson, R. P.,
	Papaloizou, J. C. B., \&
	Torquem, C.,
	1996,
	\mnras, 282, 597

\bibitem[Larwood(1998)]{larwood98}
	Larwood, J. D.,
	1998,
	\mnras, 299, L32

\bibitem[Lee, Saio \& Osaki(1991)]{lee91}
	Lee, U.,
	Saio, H., \&
	Osaki, V.,
	1991,
	\mnras, 250, 432

\bibitem[Lyubimkov et al.(2002)]{lyubimkov02}
	Lyubimkov, L. S.,
	Rachkovskaya, T. M.,
	Rostopchin, S. I., \&
	Lambert, D. L.,
	2002,
	\mnras, 333, 9

\bibitem[Madore \& Freedman(1991)]{madore91}
	Madore, B., \& 
	Freedman, W., 1991, 
	PASP, 103, 667

\bibitem[Melchior, Hughes \& Guibert(2000)]{melchior00}
	Melchior, A. L.,
	Hughes, S. M. G.,
	Guibert, J., 2000,
	A\&AS, 145, 11

\bibitem[Mennickent \& Sterken(1997)]{mennickent97}
	Mennickent, R.E., 
	\& Sterken, C., 
	1997, 
	\aap, 121, 113

\bibitem[Millar \& Marlborough(1999)]{millar99}
	Millar, C. E., \&
	Marlborough, J. M.,
	1999,
	\apj, 526, 400

\bibitem[Milone, Stagg \& Schiller(1992)]{milone92}
	Milone, E. F.,
	Stagg, C. R., \&
	Schiller, S. J., 
	1992, 
	in Evolutionary Processes in Interacting Binary Stars,
	ed. Kondo, Y.
	(Dordrecht; Boston: Kluwer), 479
	
\bibitem[Monnier \& Millan-Gabet(2002)]{monnier02}
	Monnier, J.D., \&
	Millan-Gabet, R.,
	2002,
	\apj, 579, 694

\bibitem[Mouillet et al.(2001)]{mouillet01}
	Mouillet, D.,
	Lagrange, A. M.,
	Augereau, J. C., \&
	M$\acute{e}$nard, F.,
	2001,
	\aap, 372, L61

\bibitem[Ozernoy et al.(2000)]{ozernoy00}
	Ozernoy, L. M.,
	Gorkavyi, N. N.,
	Mather, J. C., \&
	Taidakova, T. A.,
	2000,
	\apj, 537, L147

\bibitem[Papaloizou \& Terquem(1995)]{papaloizou95}
	Papaloizou, J. C. B., \&
	Terquem, C.,
	1995,
	\mnras, 274, 987

\bibitem[Polidan \& Peters(1976)]{polidan76}
	Polidan, R. S., \&
	Peters, G. J.,
	1976,
	in IAU Sym. 70, Be and Shell Stars,
	ed. Slettebak, A.,
	(Boston: D. Reidel), 59

\bibitem[Press, Flannery, Teukolsky \& Vetterling(1992)]{press92}
	Press, W.,
	Flannery, B.,
	Teukolsky, S., \&
	Vetterling, W.,
	1992,
	Numerical Recipies in C,
	(Cambridge: Cambridge U. Press)
 
\bibitem[Quirrenbach et al.(1997)]{quirrenbach97}
	Quirrenbach, A., et al., 
	1997,
	\apj, 479, 477

\bibitem[Sandberg-Lacy, Helt \& Vaz(1999)]{sandberg99}
	Sandberg Lacy, C.H.,
	Helt, B.E., \&
	Vaz, L.P.R., 
	1999,
	\aj, 117, 541

\bibitem[Soderhjelm(1974)]{soder74}
	Soderhjelm, S., 
	1974,
	Informational Bulletin on Variable Stars \# 885

\bibitem[Tokunaga et al.(2004)]{tokunaga04}
	Tokunaga, A.T., et al.
	2004,
	\apj, 601, L91

\bibitem[Torres \& Stefanik(2000)]{torres00}
	Torres, G., 
	\& Stefanik, R.P., 
	2000,
	\aj, 119, 1914

\bibitem[Udalski, Kubiak \& Szymanski(1997)]{udalski97}
	Udalski, A., 
	Kubiak, M.,\& 
	Szymanski, M.,
	1997,
	\actaa 47, 319 (OGLE-II)

\bibitem[Udalski(2003)]{udalski03}
	Udalski, A.,
	2003,
	\actaa 53, 291 (OGLE-III)

\bibitem[Weinberger et al.(2000)]{weinberger00}
	Weinberger, A. J.,
	Rich, R. M.,
	Becklin, E. E.,
	Zuckerman, B., \&
	Matthews, K,
	2000,
	\apj, 544, 937

\bibitem[Winn et al.(2003)]{winn2003}
	Winn, J.N.,
	Garnavich, P.M.,
	Stanek, K.Z., \& 
	Sasselov, D.D., 
	2003,
	\apj, 593, L121

\bibitem[Winn et al.(2004)]{winn04}
	Winn, J.N.,
	Holman, M.J.,
	Johnson J.A.,
	Stanek, K.Z., \& 
	Garnavich, P.M., 
	2004,
	\apj, 603, L45

\bibitem[Wyatt et al.(1999)]{wyatt99}
	Wyatt, M. C.,
	Dermott, S. F.,
	Telesco, C. M.,
	Fisher, R. S.,
	Grogan, K.,
	Holmes, E. K., \&
	Pi$\tilde{n}$, R. K.,
	1999,
	\apj, 527, 918

\bibitem[Wyrzykowski et al.(2003)]{wyr03}
	Wyrzykowski, \L., et al.
	2003, 
	\actaa, 53, 1

\bibitem[Zebru\'n et al.(2001)]{zebrun01}
	\.{Z}ebru\'n, K., et al.
	2001, 
	\actaa, 51, 317

\end{thebibliography}
\end{document}